%% file: main_JM.tex
\documentclass[final,3p,times,twocolumn,authoryear]{elsarticle}
\usepackage{amssymb}
\usepackage{amsthm}
\usepackage[squaren]{SIunits}
\usepackage{xspace}
\usepackage{bm}
\usepackage{multirow}

\usepackage[table]{xcolor}

\journal{Journal of Colloid and Interface Science}

\include{macros}

\begin{document}

\begin{frontmatter}


\title{Spontaneous spinning of a dichloromethane drop\\ on \color{black}{an aqueous} \color{black} surfactant solution}

\author[inst1]{Dolachai Boniface\corref{cor1}}
\cortext[cor1]{Corresponding author}
\ead{dolachai.boniface.postdoc@gmail.com}

\affiliation[inst1]{organization={F\'ed\'eration de Recherche FERMAT},
            addressline={Universit\'e de Toulouse, CNRS}, 
            city={Toulouse},
            country={France}}\, 
\author[inst1,inst2]{Julien Sebilleau\corref{cor2}}
\cortext[cor2]{Corresponding author}
\ead{julien.sebilleau@imft.fr}

\affiliation[inst2]{organization={Institut de M\'ecanique des Fluides de Toulouse (IMFT)},
            addressline={Universit\'e de Toulouse, CNRS}, 
            city={Toulouse},
            country={France}}
            \author[inst1,inst2]{Jacques Magnaudet}
            \ead{jmagnaud@imft.fr}
            \author[inst1,inst3]{V\'eronique Pimienta}
             \ead{veronique.pimienta@univ-tlse3.fr}

\affiliation[inst3]{organization={Laboratoire des IMRCP, Universit\'e de Toulouse, CNRS UMR 5623},
            addressline={Universit\'e Toulouse III - Paul Sabatier}, 
            city={Toulouse},
            country={France}}          
         
\begin{abstract}
\textcolor{black}{
We report a series of experiments carried out with a dichloromethane drop deposited on the surface of an aqueous solution containing a surfactant, cetyltrimethylammonium bromide. After an induction stage during which the drop stays axisymmetric, oscillations occur along the contact line. These oscillations are succeeded by a spectacular spontaneous spinning of the drop. The latter quickly takes the form of a two-tip `rotor' and the spinning rate stabilizes at a constant value, no longer varying despite the gradual changes of the drop shape and size. The drop eventually disappears due to the continual dissolution and evaporation of dichloromethane. \\
\indent Schlieren visualizations and particle image velocimetry are used to establish a consistent scenario capable of explaining the evolution of the system. The Marangoni effect induced by the dissolution of dichloromethane in the drop vicinity is shown to be responsible for the observed dynamics. Arguments borrowed from dynamical systems theory and from an existing low-order model allow us to explain qualitatively why the system selects the spinning configuration. The geometry of the immersed part of the drop is shown to play a crucial role in this selection process, as well as in the regulation of the spinning rate. }

\end{abstract}


\begin{graphicalabstract}
    \centering
    \includegraphics[width=0.89\textwidth]{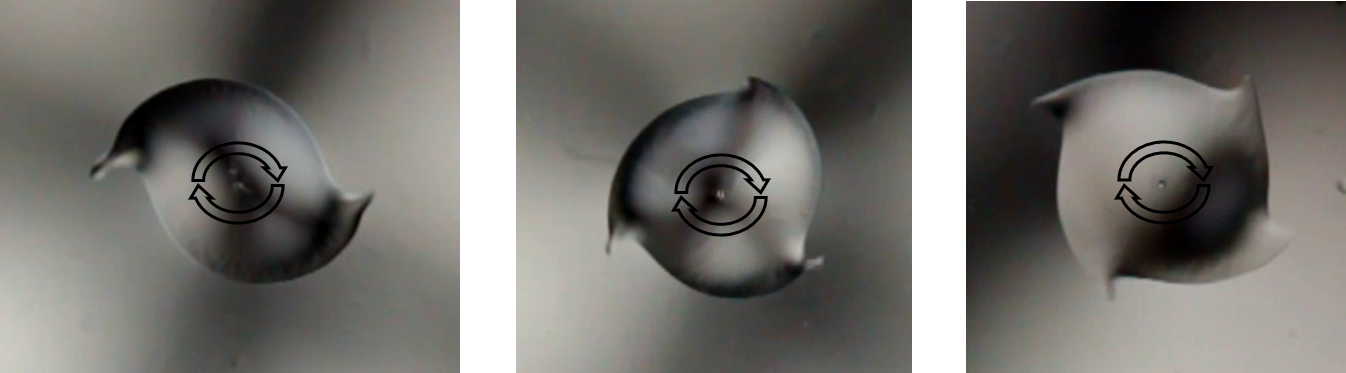}
\end{graphicalabstract}



\end{frontmatter}


\section{Introduction}

`Active' drops are simple fluid systems, consisting most often of two immiscible or partially miscible fluids and a surfactant, that exhibit remarkably rich and diverse dynamics, such as spontaneous translation ~\cite{Nagai2005,Chen2009,Izri2014,Maass2016,Reichert2021}, rotation \cite{Wang2021}, repeated pulsations~\cite{Stocker2007,Chen2017,Wodlei2018}, emission of satellite droplets~\cite{Stocker2007,Keiser2017,Wodlei2018}, etc. In all cases, some mass transfer takes place, owing to diffusion, advection, evaporation or dissolution, so that the drop modifies the local properties of its surroundings, especially the interfacial tensions involved. 
For instance, nonuniform evaporation of a volatile drop component results in a temperature gradient along its surface, while dissolution of a component generates a chemical gradient. In both cases, a Marangoni effect takes place, owing to the non-uniformity of interfacial tensions \cite{Nakata2019}.  This effect is at the core of the amazing dynamical behaviors exhibited by such far-from-equilibrium systems. Motivated by potential applications in areas such as chemical microreactors, drug delivery, inkjet printing or  
self-cleaning of surfaces, there is a growing interest in self-propelled drops, since they allow chemical components to be transported from one place to another without requiring any external mechanical action (see \cite{Lohse2020} for a review). 
Active drops may be `swimming', i.e. propelling in the bulk of the second liquid \cite{Maass2016}, or `surfing', i.e. propelling at its surface. A swimming drop is stabilized by a surfactant layer which becomes inhomogeneous due to a dissolution process or a chemical reaction. Thus a Marangoni stress is induced at the drop surface, leading to its movement~\cite{Izri2014,Maass2016}. In the case of surfing drops, the Marangoni stress may be induced by evaporation~\cite{Reichert2021}, the combined effects of dissolution and evaporation~\cite{Nagai2005,Chen2009}, or a combination of dissolution, evaporation and surfactant migration~\cite{Izri2014}. Related phenomena such as drop pulsations \cite{Chen2017}, spontaneous rotation or emission of satellite droplets \cite{Keiser2017} are also directly linked to Marangoni effects induced by evaporation, dissolution, surfactant transport or chemical reactions in specific fluid systems.\\
\indent The system on which the present study focuses involves three components, namely a dichloromethane (DCM) drop released on a bath of water containing cetyltrimethylammonium bromide (CTAB), a surfactant, at a specific concentration. Previous studies performed with the same three components ~\cite{Pimienta2011,Pimienta2014,Antoine2016,Wodlei2018} revealed that changing the CTAB concentration may result in dramatically different dynamical behaviors. In the absence of CTAB, DCM totally wets water, leading to a fast spreading and a quick break-up of the drop at the water surface.  Adding a minute amount of CTAB in water stabilizes the drop, as it lowers the air/water surface tension, thus reducing the spreading parameter of the system. Under such conditions, right after having been deposited, the drop takes a classical circular lens shape. This shape is preserved during an induction phase beyond which the drop becomes active, developing a dynamical behavior that depends on the CTAB concentration. On increasing this concentration, the drop is successively observed to self-propel erratically while emitting continuously a string of droplets (with $0.25\,$mM CTAB), pulsate and generate periodically satellite droplets ($0.5\,$mM), take a characteristic S-shape and spin regularly ($10\,$mM), or exhibit a polygonal shape with numerous tips along the contact line ($30\,$mM). In what follows, we shall focus on the penultimate case where the DCM drop develops a spectacular spinning motion.\\
\indent In most studies dealing with spinning drops, the rotation results from an external forcing, such as an electromagnetic field for magnetic droplets \cite{Liao2017}, or an imposed temperature gradient~\cite{Basu2008}. 
Few examples of spontaneously spinning drops have been reported so far. One of such systems involves a sulfuric acid drop deposited on a liquid metal substrate made of a mixture of indium and gallium \cite{Wang2021}. 
A chemical reaction between the acid and the bath takes place when the drop is released, generating a ring of microbubbles along the contact line. After some time, the drop develops an asymmetric comma-like shape and starts rotating. The rotation stabilizes at a rate of about 2 revolutions per second and lasts for several minutes. 
Another system involves a polymer drop made of polyvinylidene fluoride dissolved in dimethylformamide \cite{Zhang2017}. As the solvent spreads over the water surface, the drop (actually a solvent-swollen gel-made object) begins to rotate and quickly reaches a rotation speed of about 5 revolutions per second, which it maintains for approximately 30 seconds. Then, contamination of the water surface by the solvent lowers the surface tension of the bath. This makes the rotation rate decrease and oscillate from 2 to 3 revolutions per second for several minutes, until the spinning stops. Obviously, the spontaneous spinning motion of both systems is rooted in a Marangoni effect. The surface tension gradient results from a chemical reaction in the former case, and from the dissolution of a chemical compound in the latter.\\
\indent The spontaneous spinning of a rigid object floating on a liquid bath is better documented~\cite{Mitsumata2001,Bassik2008,Pimienta2014,Maggi2015,Dorbolo2016}.
In most cases, the body releases a surfactant \cite{Mitsumata2001,Bassik2008} or acts as a heat source~\cite{Maggi2015}. A chiral asymmetry is introduced in the system either through the design of the object (imposing a ratchet-gear shape or a `S' shape for instance \cite{Bassik2008,Pimienta2014}), or through the position of the surfactant/heat source \cite{Maggi2015}, leading to a nonzero torque inducing the spinning motion. However, in some cases a spontaneous symmetry breaking may happen despite the lack of chiral asymmetry of the object. For instance, it has been reported that a cubic body made of a swollen amphiphilic polymer gel may spin for several minutes with a rotation rate of 3 to 6 revolutions per second \cite{Mitsumata2001}. In this case, the Marangoni effect results from the osmotic pumping of the organic solvent from the gel and its spreading on the water surface.
  A circular ice disk melting at the surface of a water pool has also been observed to spin spontaneously. In this case, the driving mechanism totally differs from those considered above, being of purely hydrodynamic origin \cite{Dorbolo2016}. Here, the cooling of the water at the base of the disk makes its density larger than that of the water located lower in the bath, generating a downward plume through the classical Rayleigh-Taylor mechanism. The purely axisymmetric flow in the plume being unstable with respect to circumferential disturbances, a vertical vortex sets in under the ice disk, putting it in rotation through viscous entrainment. \\  
  \indent The aforementioned two mechanisms, namely a Marangoni effect at the bath surface or a downward vortex flow beneath the drop, are potential candidates to explain the self-spinning behavior reported in \cite{Pimienta2011} for a DCM drop on a $10\,$mM CTAB aqueous solution. Indeed, the well-known volatility of DCM under standard conditions combined with the  S-shape exhibited by the drop during its spinning make the situation look consistent with the existence of a Marangoni torque. On the other hand, DCM being heavier than water and partially soluble in it, the presence of a vertical vortex in the bath, similar to that observed in the case of the melting ice disk, would not be unlikely. 
The aim of this paper is to provide a detailed analysis of the dynamics of the DCM drop under such conditions, and to elucidate the underlying driving mechanisms. The paper is organized as follows. The experimental setup and the measurement techniques are presented in section 2. Section 3 makes use of the optical observations performed with a Schlieren setup to describe the evolution of the drop, from its deposit at the surface of the bath to its disappearance through dissolution and evaporation. In section 4, we analyze the time variations of the spatial structure of the flow in the aqueous bath, as revealed by particle image velocimetry (PIV) measurements. Making use of the data provided by these two complementary techniques, we devote section 5 to reviewing and discussing critically the various physical mechanisms at work, in order to set up a consistent scenario capable of explaining the remarkable spontaneous drop spinning exhibited by this simple system. 

\section{Materials and Methods}

The experimental setup consists of a $5\times5\times5\,$cm$^{3}$ optical glass container filled with a $10\,$mM CTAB aqueous solution. As the container dimensions are precisely known, the depth of the aqueous solution is adapted by tuning the volume poured inside. The DCM drop (99.9\% Sigma-Aldrich) is obtained by depositing successively smaller drops, using a gas tight syringe. The total volume injected ranges from 5 to 40 $\mu$L; most of experiments were performed with a 20$\mu$L drop. The aqueous CTAB solution is obtained by dissolving CTAB crystal powder in Millipore-ultrapure water. The solubility of CTAB being close to the critical micellar concentration ($1\,$mM) at room temperature ($\simeq$ 20$^{\circ}$C), the aqueous solution is heated to reach the desired concentration of $10\,$mM. Then the solution is cooled down to room temperature and remains crystal-free (hence usable for experiments) for a few hours. CTAB being hydrophobic, its partition is strongly favorable to the organic phase. CTAB also shows a strong affinity with both water/air and water/DCM interfaces \cite{Tadmouri2008}. \textcolor{black}{In particular, measurements performed with the pendent drop technique indicate that the surface tension of the $10\,$mM CTAB solution, $\gamma_{w/a}$, is $36\pm1$ mN.m$^{-1}$ instead of $72.8\,$mN.m$^{-1}$ for pure water. The same technique gives the DCM surface tension as $\gamma_{o/a}=28.2\pm0.3$ mN.m$^{-1}$} (see figure 1 in the Supplemental Material (SM)). 
It must also be noted that DCM is highly volatile (boiling at $39.6^{\circ}$C), denser than water ($\rho_o=1.32\,$g.cm$^{-3}$) and partially soluble in it (solubility = $0.15\,$M.L$^{-1}$). 
\subsection{Schlieren optical setup}
\begin{figure}[h!]
    \centering
    \includegraphics[width=0.45\textwidth]{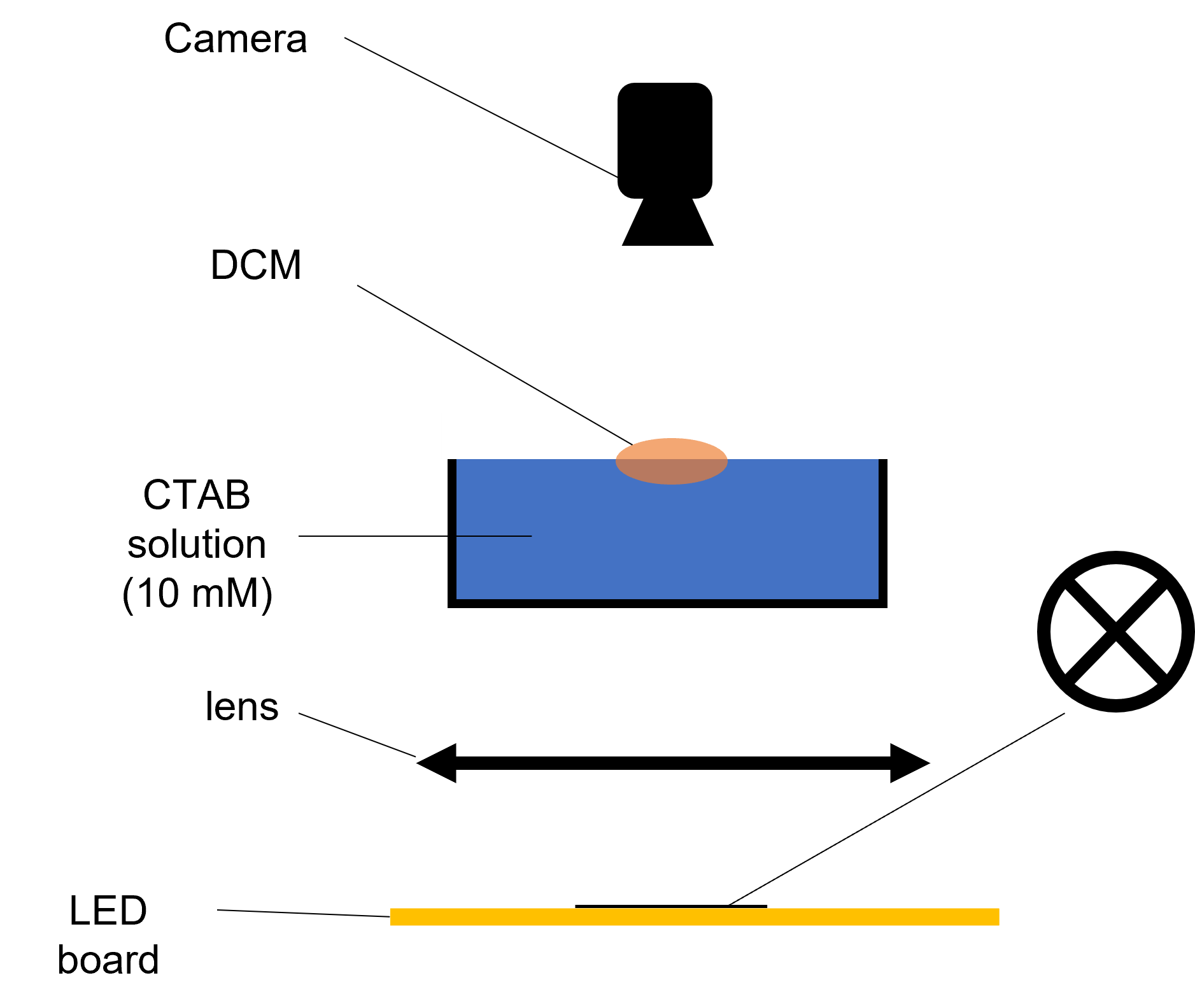}
       \vspace{-4mm}
        \caption{Schlieren optical setup}
    \label{fig:Schlieren}
\end{figure}

Two different optical diagnostics were used in the course of this study. First, a Schlieren method allowed us to follow the evolution of the drop shape throughout its lifetime. Second, a particle image velocimetry (PIV) equipment made it possible to determine the flow field in the aqueous bath.\\
\indent As both the aqueous CTAB solution and the DCM drop are transparent, direct observation of the system is difficult. To obtain images with a sufficient contrast, one can take advantage of the difference in the refractive indices of the two liquids ($n_{water}=$1.3 and $n_{DCM}=$1.4) by using the Schlieren setup sketched in figure~\ref{fig:Schlieren}. 
At the bottom of the setup stands a LED backlight with a black paper cross laid on it. The light emitted by the LED board goes through a convex lens with a $12\,$cm focal length, and then crosses the water pool and the DCM drop. The resulting image is then recorded by a camera (Canon 600D - 50 fps /Andor Zyla 5.5 - 100 fps) with a spatial resolution of $20.8/19.1\,$pix.mm$^{-1}$. The vertical distance between the lens and the bottom of the container is adjusted so as to obtain images whose contrast is sufficient to determine the drop shape after a suitable post-processing (see figure~\ref{fig:induction_side}$(a)$). In addition to these Schlieren visualizations, classical views from the edge allow us to record the submerged surface of the drop (i.e. the part of its surface in contact with the aqueous solution), the upper surface (i.e. the part in contact with air) being hidden by the meniscus made by the aqueous solution with the container walls (see the inset in figure~\ref{fig:induction_side}$(c)$). Image processing used with both the Schlieren visualizations and the edge views is based on a homemade Python program making use of the graphic libraries OpenCV and Skimage. Post-processed images give access to the instantaneous position and orientation of the horizontal drop contour, and to the vertical projection of the drop-bath interface.

\subsection{PIV setup}
\begin{figure}
    \centering
    \includegraphics[width=0.45\textwidth]{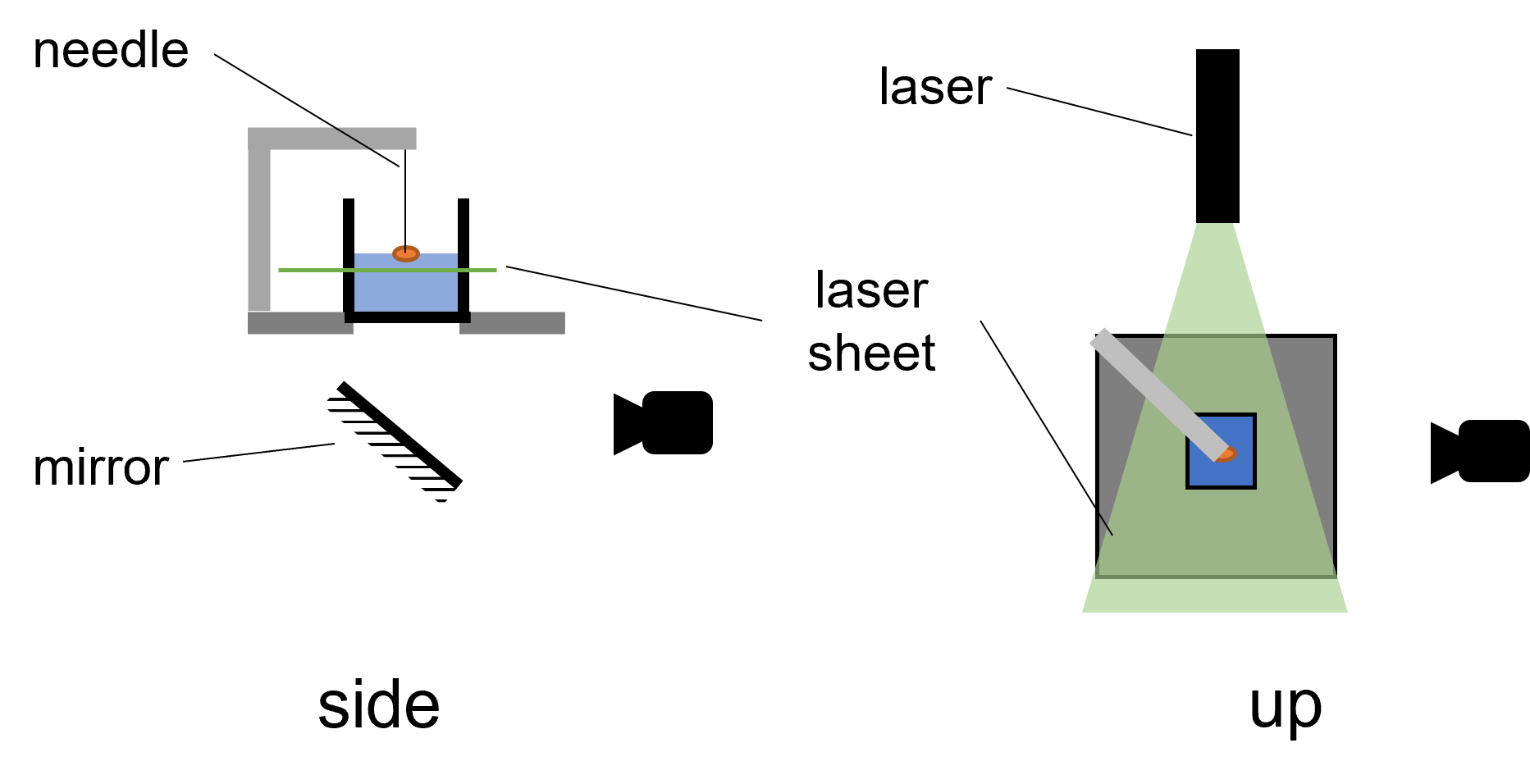}
       \vspace{-2mm}
        \caption{PIV setup (horizontal configuration)}
    \label{fig:PIV_setup}
\end{figure}

To complement the Schlieren visualizations, PIV measurements are carried out to investigate the flow in the aqueous solution. In these runs, the drop is pinned with a needle located right at the center of the tank to maintain it fixed and avoid possible erratic horizontal motions. The whole device is installed on an air cushion table (optical smart table ST-UT2) to minimize vibrations. A continuous laser (S3 Krypton Series-class 4-power $750\,$mW) with a $532\,$nm wavelength is employed to create a thin laser sheet ($150\,$nm thick) that illuminates a horizontal or vertical plane of the pool. The aqueous bath is seeded with polyamide PA12 spheres having a mean diameter of $7.25\,\mu$m with Rhodamine molecules adsorbed at the surface. As Rhodamine is fluorescent, with a maximum light absorption at $530\,$nm and a maximum emission at $590\,$nm, an optical high-pass filter is used. In this way, any undesired reflection of the laser sheet at the drop/bath interface is removed and only tracers are visible on the images. Images are recorded thanks to a high-speed camera (DIMAX S4) operating at $400\,$fps with a spatial resolution of $24.4\,$pix.mm$^{-1}$. A vertical laser sheet passing through the drop center is created to obtain velocity fields in the vertical midplane of the bath. As sketched in figure~\ref{fig:PIV_setup}, a horizontal laser sheet is produced by inserting a $45^{\circ}$-mirror below the glass container to obtain velocity fields in horizontal planes located at various depths below the drop (this configuration represents the majority of the runs).
The recorded images are processed with the DAVIS software. \textcolor{black}{Images are cropped and first analyzed roughly over $128\times128$ pixels windows with a $50 \%$ overlap. Then, four extra analyses are performed over $16\times16$ pixels windows with a $50 \%$ overlap.} 
This procedure gives access to the two-dimensional velocity fields, from which kinematic quantities involving the in-plane velocity gradients can be computed (see section 4). Note that, since DCM dissolves into water and has a different refractive index, velocity vectors determined near the drop surface may be incorrect. For this reason, special care was taken in the analysis of this region and outlier vectors were removed. Particles used as tracers need to be stabilized with a surfactant to avoid agglomeration, a role that CTAB fulfills properly. To ensure that tracers do not affect the behavior of the spinning drop, we performed Schlieren observations with the seeded CTAB aqueous solution prior to the PIV measurements. These observations did not show any significant influence of the tracers \textcolor{black}{(see figure 3 in SM)}. 

\section{Evolution of the drop shape}
\begin{figure}[h!]
    \centering
    \includegraphics[width=0.49\textwidth]{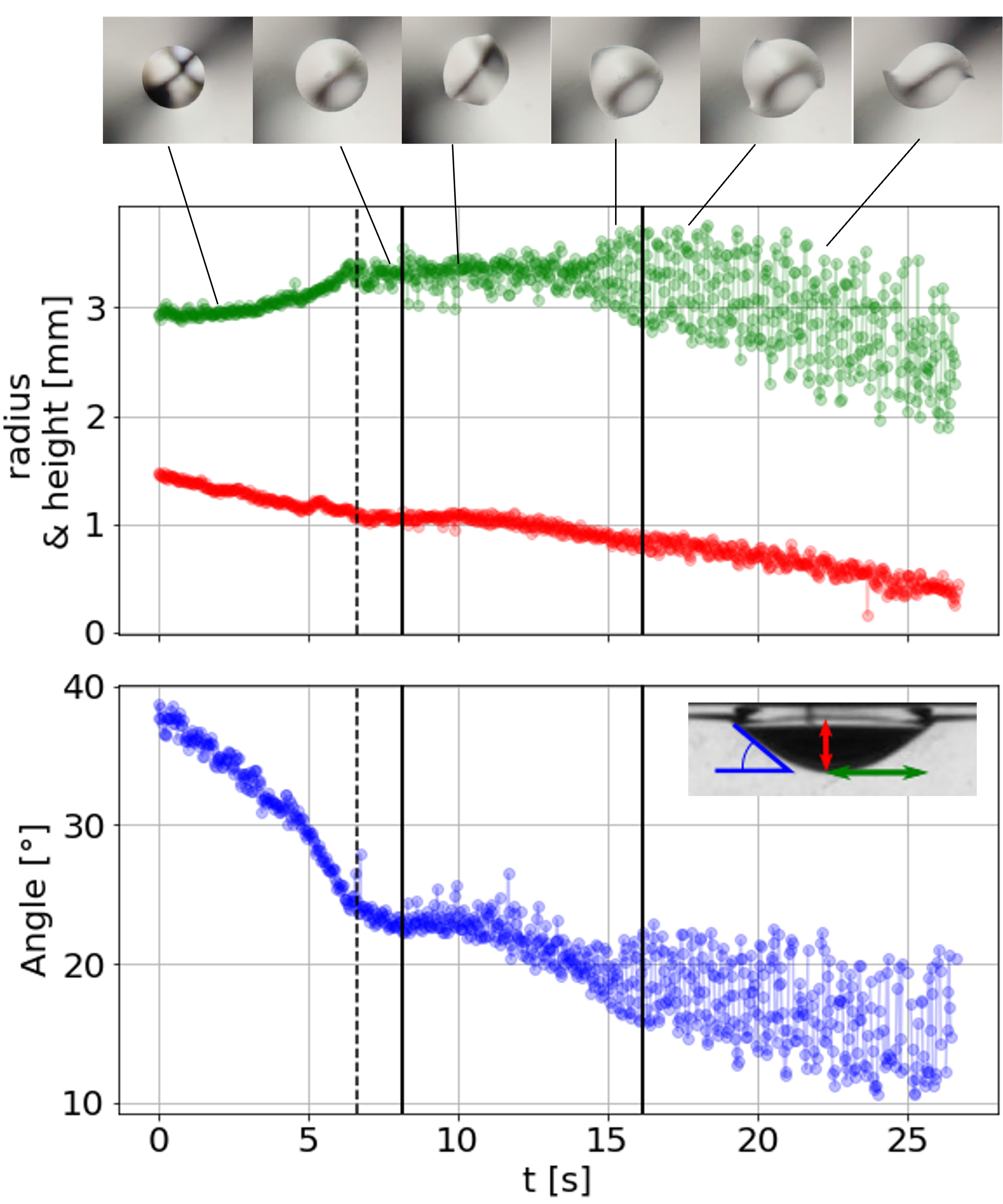}
    \caption{Evolution of the drop shape and size during a typical run. $(a)$: top view of the drop at six different characteristic instants of time; $(b)$: apparent radius (green symbols) and submerged height (red symbols); $(c)$: apparent contact angle $\theta_{w/o}$ between the drop and the aqueous solution (blue symbols); see the inset in $(c)$ for the definition of all three quantities. The two vertical solid lines in $(b)-(c)$ mark the approximate frontiers between the induction, vibration and spinning stages; the dotted line indicates the time by which the system reaches approximately time-independent characteristics before vibrations start. 
    }
    \vspace{-125mm}\hspace{-73mm}$(a)$\\
      \vspace{35mm}\hspace{-73mm}$(b)$\\
       \vspace{34mm}\hspace{-73mm}$(c)$\\
    \vspace{41mm}
    \label{fig:induction_side}
\end{figure}

A typical evolution of the drop shape determined with the Schlieren technique is illustrated in figure~\ref{fig:induction_side}. Three distinct stages may easily be identified. During a first `induction' stage, the drop adopts a classical lens shape with a circular contour and spreads a little on the bath surface. Then, during a short transient, waves arise along the contour, yielding a `vibration' stage. Finally, the drop adopts a helicoidal S-shape and spins regularly while ejecting radially small droplets. These are the characteristics typical of the spinning stage. During the entire process, evaporation, dissolution and droplet ejection lower gradually the volume of the mother drop, until it eventually disappears (\textcolor{black}{(see video 1 in SM)}.

\subsection{Induction stage}
The induction stage starts just after the drop is deposited. \textcolor{black}{In the example displayed in figure~\ref{fig:induction_side}, this stage lasts for approximatively $8\,$s. However this duration varies strongly from one run to another, due especially to the aging of the water/drop and water/air interfaces during the deposition process, and also to the ambient temperature. A typical duration of this stage is $15\,$s, but durations up to $70\,$s have been observed in some runs.} 
As soon as the drop is deposited, DCM starts to dissolve into the bath, which alters the physicochemical properties of the air/solution interface. By inserting a Wilhelmy plate `far' from the drop \textcolor{black}{(see figure 2$(a)$ in SM)}, we determined that the surface tension of the aqueous solution decreases abruptly by \textcolor{black}{$1\pm0.2\,$mN.m$^{-1}$} from the time the drop first touches the surface to the moment it is completely deposited. During the induction stage, dissolution goes on but its effect on the surface tension slows down, so that the surface tension decreases typically by another $0.5\,$mN.m$^{-1}$ from the end of the deposition to the end of the induction stage (figure 2$(b)$ in SM). Hence, at the end of the deposition ($t=0\,$s in figure \ref{fig:induction_side}), the surface tension of the solution \textit{far} from the drop, say $\gamma_{w/a}^\infty$,  is close to \textcolor{black}{$35\,$}mN.m$^{-1}$ instead of $36\,$mN.m$^{-1}$, while at the end of the induction stage ($t\approx8\,$s in figure \ref{fig:induction_side}) $\gamma_{w/a}^\infty$ is close to $34.5\,$mN.m$^{-1}$.

During the induction stage, the drop slightly spreads over the aqueous solution and takes a classical lens shape. 
As figures \ref{fig:induction_side}$(b)-(c)$ show, the spreading process results in a slight increase of the radius and a decrease of both the submerged height and the apparent contact angle. Meanwhile, the drop centroid exhibits some erratic motions, with an amplitude less than one drop diameter.  A  `quasi-steady' shape resulting from a balance between buoyancy and capillary forces~\citep{DeGennes2013} is reached after \textcolor{black}{approximately $7\,$s (on average, this quasi-steady state is rather reached after $10\,$s).} 
\textcolor{black}{The buoyancy force slowly decreases over time, owing to the continual evaporation and dissolution of DCM. Meanwhile, the net capillary force acting along the contact line varies due to CTAB adsorption at the DCM/aqueous solution interface. This adsorption mechanism makes the corresponding interfacial tension, $\gamma_{w/o}$, decrease gradually over time. This process typically lowers $\gamma_{w/o}$ by $4\,$mN.m$^{-1}$ over a $100\,$s period \textcolor{black}{(see figure 1 in SM)}. For this reason,  $\gamma_{w/o}$ decreases from $7\,$mN.m$^{-1}$ by the time the drop is deposited to $6\,$mN.m$^{-1}$ at the end of the induction stage. As figure \ref{fig:induction_side}$(c)$ indicates, the drop/solution contact angle, $\theta_{w/o}$, decreases from an initial value close to $38^{\circ}$ to approximately $23^{\circ}$ at the end of the induction stage. From $t\approx7\,$s to $t\approx10\,$s, the various geometric indicators displayed in figures \ref{fig:induction_side}$(b)-(c)$ suggest that the system has reached a quasi-steady configuration, with an apparent radius $R_0\approx3.4\,$mm and a submerged height $e_0\approx1\,$mm. Considering $\gamma_{o/a}=28\,$mN.m$^{-1}$, $\gamma_{w/o}=6\,$mN.m$^{-1}$ and $\theta_{w/o}=23^\circ$, equilibrium conditions at the contact line \cite{DeGennes2013} imply $\gamma_{w/a}\approx33.5\,$mN.m$^{-1}$ there (see section 2 in SM). This estimate\footnote{\textcolor{black}{Although the DCM/air contact angle is unknown, it is clearly small. Varying this angle by $3^\circ$ only changes the prediction for $\gamma_{w/a}$ by $0.1\,$mN.m$^{-1}$, making it possible to show that the above estimate for $\gamma_{w/a}$ is as accurate as that for $\gamma_{w/o}$ (see section 2 in SM).}} reveals that $\gamma_{w/a}$ is reduced by \textcolor{black}{approximately $1\,$}mN.m$^{-1}$ \textit{at the contact line} compared to its `far-field' value $\gamma_{w/a}^\infty\approx34.5\,$mN.m$^{-1}$. This is not unlikely since DCM dissolution is expected to be stronger close to the contact line, increasing locally the DCM concentration in the bath, which inescapably decreases $\gamma_{w/a}$. Of course, the direct consequence of this surface tension difference between the `far field' and the contact line, say $\Delta\gamma_{w/a}$, is that a Marangoni stress tending to move the surface of the bath away from the drop is present at the end of the induction stage. With the above estimates, the spreading parameter $S=\gamma_{w/a}-(\gamma_{o/a}+\gamma_{w/o})$ at the contact line is approximately $-0.5\pm0.2\,$mN.m$^{-1}$ at the end of the induction stage (section 2 in SM). Interestingly, with $\theta_{w/o}=38^\circ$ and $\gamma_{w/o}=7\,$mN.m$^{-1}$ right after the deposition is completed, the initial spreading parameter is found to be $S\approx-1.2\pm0.2\,$mN.m$^{-1}$. The above two estimates indicate that, although it remains negative throughout the induction stage, the spreading parameter decreases over time (in absolute value), which is consistent with the slight spreading of the drop noticed in figure \ref{fig:induction_side}$(b)$.  
 The tiny equilibrium value of $S$ at the end of the induction stage indicates that the drop `hesitates' to recoil or spread out and prefigures the occurrence of oscillations along the contact line.} 

\subsection{Vibration stage}
\label{32}
\begin{figure}[h!]
\vspace{1mm}
    \centering
    \includegraphics[width=0.48\textwidth]{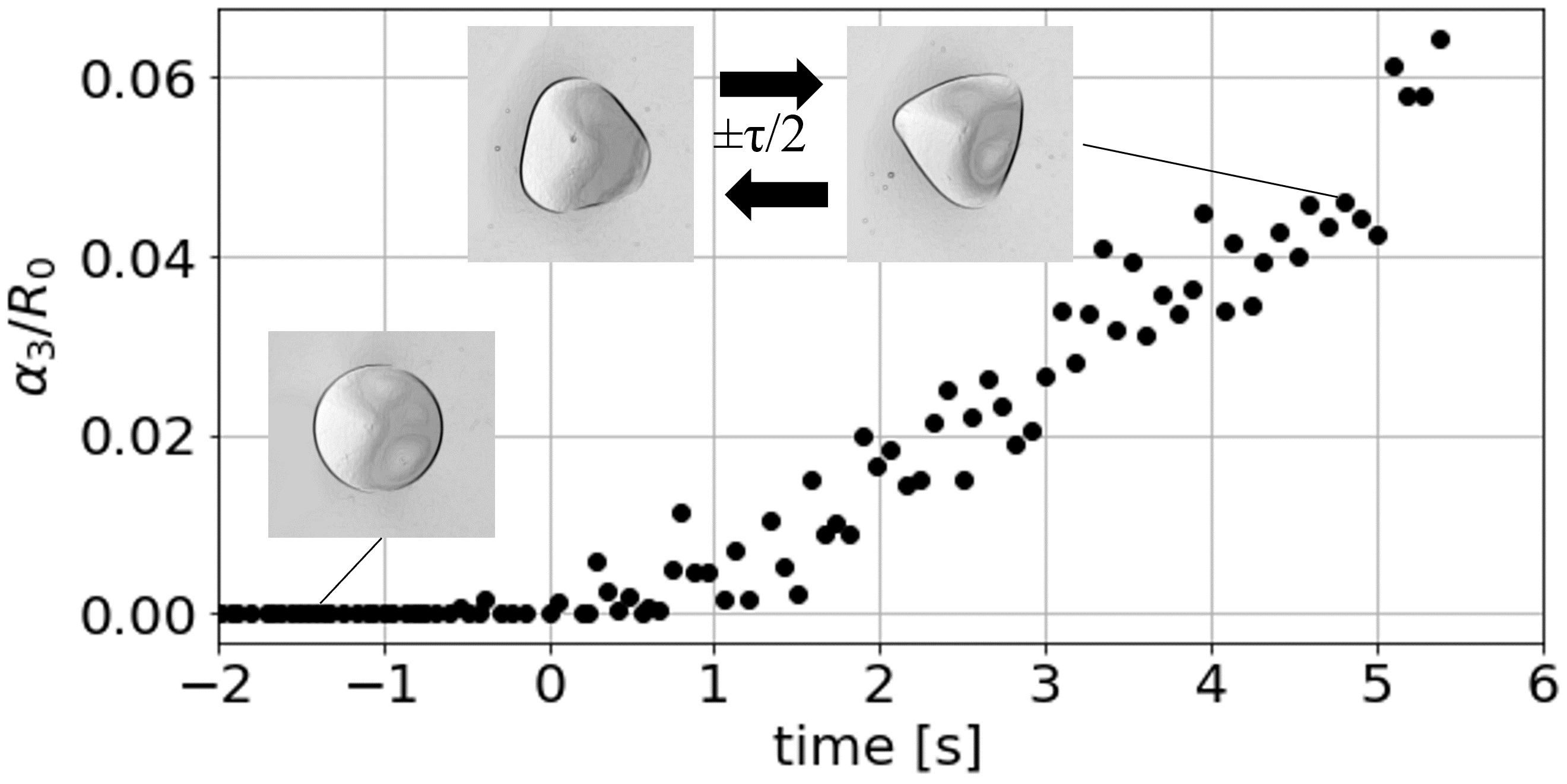}
    \caption{Amplitude $\alpha_3$ of stationary oscillations corresponding to the mode $m=3$, normalized by the equilibrium drop radius $R_0$. The snapshots show the shape of the drop contour at different times, two of them illustrating the `triangular' shape, just before the drop starts spinning. The two snapshots are separated by half a period $\tau/2$, with $\tau=1/f_3$, $f_3$ being the associated oscillation frequency.}
    \label{fig:oscillation_r3}
    \vspace{-1mm}
\end{figure}

In most experimental runs, oscillations of the contact line are observed beyond the brief equilibrium period that concludes the induction stage. \textcolor{black}{In the majority of cases ($\approx75\%$ of the runs), these oscillations have a small amplitude and rotate along the contact line (video 3 in SM). In contrast, stationary oscillations with a significantly larger amplitude are observed in approximately $15\%$ of the runs (video 4 in SM). In the last $10\%$,} more intense events such as pulsations or `effervescence' with ejection of multiple droplets happen. This is the complex dynamical context in which the drop spontaneously loses its axial symmetry. This step lasts between $5$ to $10\,$s before the drop starts spinning. This variety of behaviors makes an extensive description of the dynamics involved during this intermediate stage difficult. For this reason we \textcolor{black}{only detail the best characterized case (although not the most frequent)}, in which the contact line exhibits stationary oscillating tips with a triangular rounded shape, as shown in the third and fourth snapshots of figure~\ref{fig:induction_side}$(a)$. Not surprisingly, expanding the drop contour in Fourier modes with respect to the polar angle reveals that the observed tips essentially correspond to the mode $m=3$ (modes with $m\geq4$ have sub-pixel amplitudes). The amplitude $\alpha_{3}$ of this mode, determined by considering the maximum displacement of the contour during each oscillation, is displayed in figure~\ref{fig:oscillation_r3}.
As this figure shows, $\alpha_3$ increases linearly over time, until the moment when the drop starts spinning and turns into a three-tip spinner (see the fifth snapshot in figure~\ref{fig:induction_side}$(a)$). Averaging all experimental runs exhibiting the $m=3$ standing mode, we found that, just before the drop starts spinning, $\alpha_{3}=140\pm 20 \,\mu$m, i.e. $\alpha_3/R_0\approx0.04$. The oscillation frequency is $f_3=6.0\pm0.3\,$Hz, and the phase switches periodically by $180^\circ$, as the two snapshots in figure \ref{fig:oscillation_r3} indicate. We are not aware of any previous observation of such an odd deformation mode of a drop resting on a liquid substrate, the closest observations being those in which the drop exhibits a pulsating behavior~\cite{Stocker2007, Chen2017}. In contrast, similar odd-mode deformations of pancake-like drops have been reported when the drop rests on a vibrated surface~\cite{Noblin2005,Shen2010, Mampallil2013}, or is supported by a gas layer induced either by a gas flow~\cite{Bouwhuis2013} or by the evaporation flux resulting from the Leidenfrost effect \cite{Ma2018}. For Leidenfrost drops, contours exhibiting up to thirteen lobes have been reported~\cite{Ma2018}. Three-lobe oscillations at the contact line similar to the present ones were also observed with pancake drops resting on a vibrated liquid substrate \cite{Noblin2004}. 
\subsection{Spinning stage}
The spinning stage succeeds the vibrating stage. This is the most spectacular part of the dynamics of the system, during which the drop is characterized by a spinning helical S-shape with two tips from which droplets are repeatedly emitted radially.
\begin{figure}[h]
    \centering
    \includegraphics[width=0.5\textwidth]{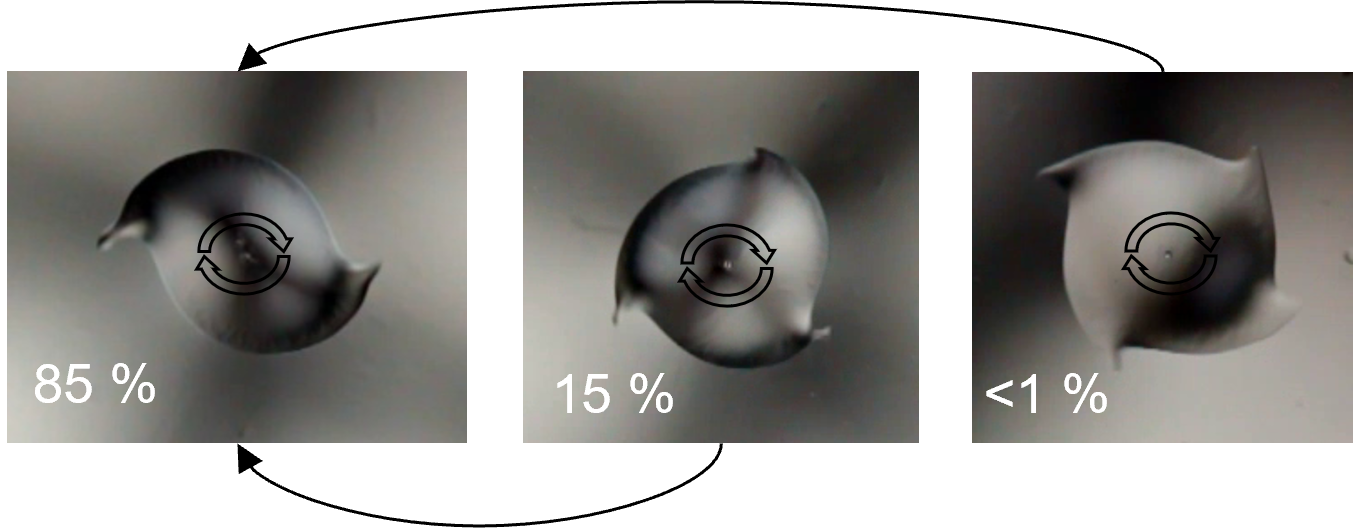}
    \caption{The various drop shapes observed at the beginning of the spinning stage (all three drops are spinning clockwise). Percentages show the proportion of occurrence of each shape; the black arrows sketch the evolution of the unstable three- and four-tip shapes.}
    \label{fig:spinners}
        \vspace{-1mm}
\end{figure}
The number of tips may vary from two to four at the beginning of this stage (figure~\ref{fig:spinners}), but the shape always evolves toward the two-tip configuration. This evolution is gradual and lasts for a few seconds. Three-tip (resp. four-tip) drops are initially observed in 15\% (resp. less than $1\%$) of the runs. However these shapes are unstable and one (resp. two) of the tips quickly fades away. The scarce four-tip configuration emerges after a `chaotic' vibration stage during which the drop contour explores numerous possible shapes ((see video 5 in SM). \textcolor{black}{The transient three-tip configuration appears to be the direct continuation of the $m=3$ stationary oscillations of the contact line observed in approximately $15\%$ of the runs during the vibration stage.} 
Not surprisingly, no preferential spinning direction emerges, i.e. we observe as many drops spinning clockwise as counterclockwise.
During its spinning motion, the drop regularly ejects droplets through the tips, while dissolving into the solution and evaporating. These three processes make it shrink gradually as the last two snapshots in figure \ref{fig:induction_side}$(a)$ illustrate. When it becomes small enough, i.e. when its apparent radius becomes typically less than $0.1\,$mm, it loses its S-shape and stops spinning. Then its behavior becomes chaotic and unpredictable until DCM dissolves and evaporates in totality, marking the disappearance of the drop.\\

\begin{figure}[t!]
    \centering
    \includegraphics[width=0.49\textwidth]{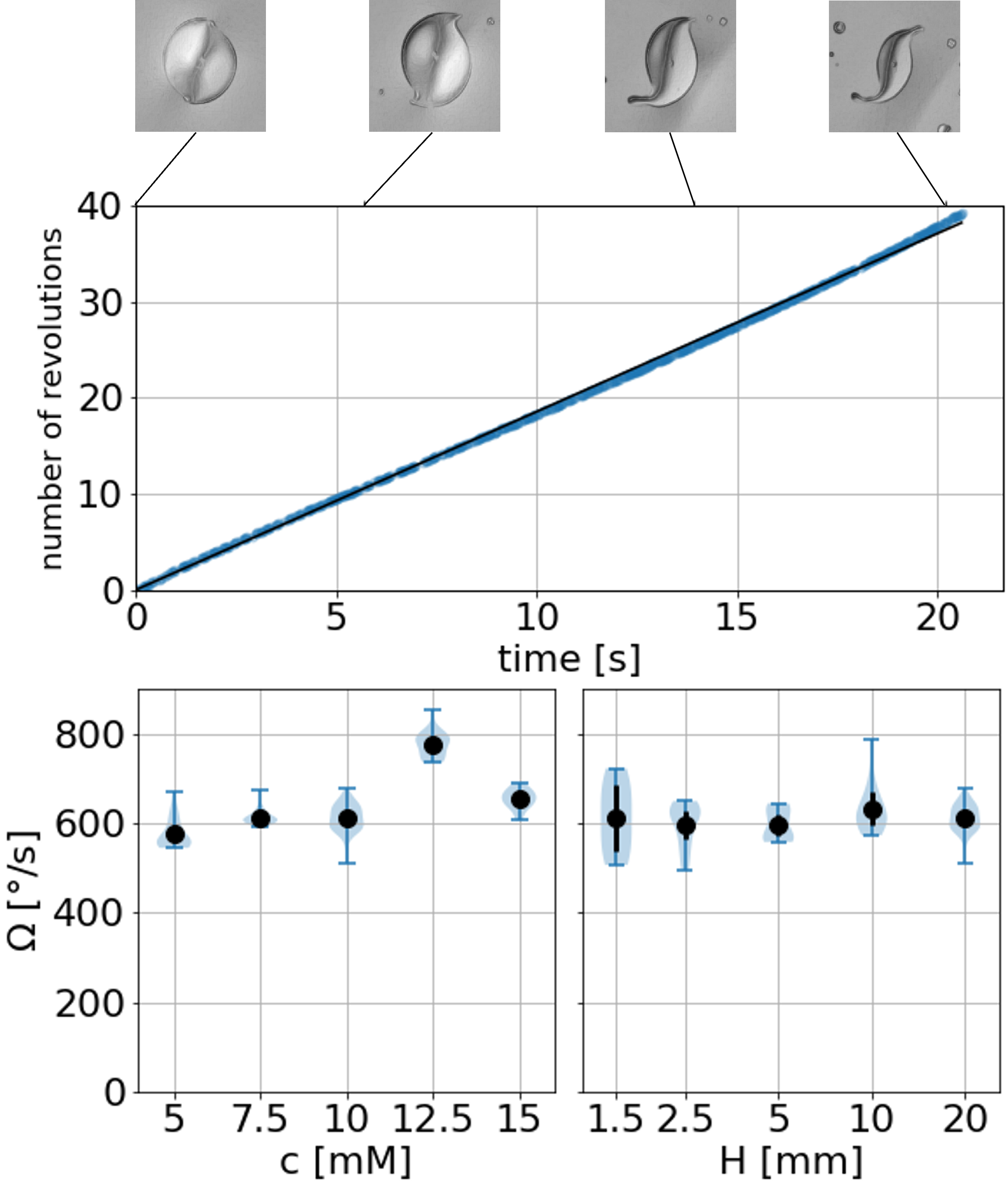}
    \caption{Characteristics of the spinning motion. $(a)$: snapshots showing the evolution of the drop shape during the spinning stage; $(b)$: number of revolutions vs. time (blue dots) during a run, and linear regression $\Omega=660^\circ$.s$^{-1}$ (solid line, $R^2=0.999$); $(c)$: variation of the spinning rate with the CTAB concentration in a $2\,$cm-deep bath 
    $(d)$: variation of the spinning rate with the depth of the bath in a $10\,$mM CTAB solution. In $(c)-(d)$, the blue area is the violin representation showing the minimum and maximum values and the probability density of the data; black dots represent the mean value, error bars correspond to a $99\%$ confidence interval obtained from Student's law.   }
     \vspace{-125.5mm}\hspace{-73mm}$(a)$\\
      \vspace{36mm}\hspace{-73mm}$(b)$\\
       \vspace{31.5mm}\hspace{-12mm}$(c)$\hspace{31.5mm}$(d)$\\
    \vspace{43mm}
    \label{fig:spinning_velocity}
\end{figure}

\emph{Spinning velocity} - 
The number of revolutions performed by the drop may be extracted by tracking the time variations of its shape during the spinning stage. Figure~\ref{fig:spinning_velocity}$(b)$ displays the corresponding result for a drop exhibiting an especially long spinning stage. The number of revolutions is seen to grow linearly over time, implying a constant rotation rate, $\Omega\approx660^\circ$s$^{-1}\approx11.5\,$rad.s$^{-1}$ in this case. This is a remarkable feature, given the mass loss and the shrinkage of the drop caused by evaporation, dissolution and ejection of DCM droplets. As figure~\ref{fig:spinning_velocity}$(a)$ reveals, the drop circumference evolves from a nearly circular shape with two small asymmetric tips to an elongated S-like shape.
That the rotation rate is independent of the drop shape is also supported by noting that the stable two-tip and transient three-tip drops both rotate at rates $\Omega=610\pm10^{\circ}$.s$^{-1}$ and $600\pm10^{\circ}$.s$^{-1}$, respectively (with a $99\%$-confidence interval in both cases). 
Figure \ref{fig:spinning_velocity}$(c)$ indicates that varying the CTAB concentration from $5$ to $15\,$mM barely affects $\Omega$, except for the peculiar case of a $12.5\,$mM concentration, in which case $\Omega$ appears to be approximately $25\%$ larger. That $\Omega$ exhibits virtually no sensitivity to the CTAB concentration over such a broad range is a noticeable feature, since the general dynamics of DCM drops in the presence of CTAB is known to be deeply influenced by the surfactant concentration \cite{Pimienta2014}. Similarly, figure \ref{fig:spinning_velocity}$(d)$ shows that varying the depth of the bath from\footnote{For a 20 \micro\liter\xspace drop, a $1.5\,$mm bath depth is the lowest limit that can be reached, since it corresponds to the initial drop height (see figure \ref{fig:induction_side}). For smaller depths, the drop fills the available space, forming a bridge between the free surface and the bottom of the tank, and starts pulsating after an induction period.}  $1.5\,$mm to $20\,$ mm leaves $\Omega$ unaffected, ruling out the possible influence of a depth-dependent confinement effect on the symmetry breaking leading to the spinning state. 
In conclusion, the spinning rate turns out to be a very stable quantity, which is essentially insensitive to the drop size, CTAB concentration and bath depth. 
\\

\begin{figure}[h!]
    \centering
    \includegraphics[width=0.45\textwidth]{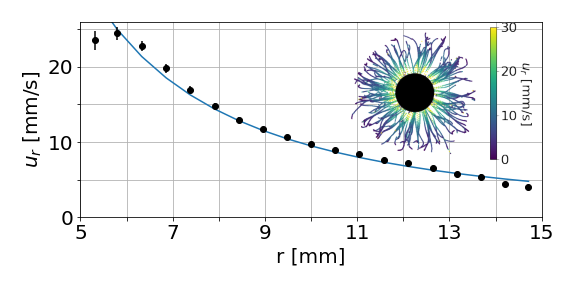}
    \vspace{-4mm}
    \caption{Radial velocity of the ejected droplets vs. the distance $r$ from the center of the mother drop. The blue line is the $u_r\sim r^{-1.75}$ power-law fit obtained through a linear regression. The inset shows the droplet paths (the dark circle in the middle is a $1\,$cm mask hiding the mother drop.}
    \label{fig:drop}
    \vspace{-5mm}
\end{figure}

\emph{Droplets ejection} -
As we saw above, droplets are regularly emitted from the drop tips and propagate radially. They generally emerge individually. However, a bigger droplet is sometimes accompanied by several smaller ones resulting from the destabilization through a Plateau-Rayleigh instability of the neck joining the droplet to the mother drop. The droplet diameters range from a few tens of $\mu$m to approximately $1\,$mm and their emission frequency at each tip is about $10\,$Hz. The droplets move radially in the vicinity of the mother drop, their speed decreasing with the radial distance. Far enough away, their trajectory is no longer strictly radial. In this late stage, they break into multiple smaller droplets. Once they have become too small to break up, dissolution and evaporation make them eventually disappear. As figure~\ref{fig:drop} indicates, the maximum droplet speed is approximately $25\,$mm.s$^{-1}$ and is reached at a radial distance $r\approx6\,$mm from the center of the mother drop. Then the radial velocity $u_r$ decreases following the approximate power law $u_r\sim r^{-1.75}$, reaching values close to $5\,$mm.s$^{-1}$ at $r\approx15\,$mm.

\section{Flow around the drop}
\label{Flow}
\begin{figure}[!t]
    \centering
      \includegraphics[width=0.49\textwidth]{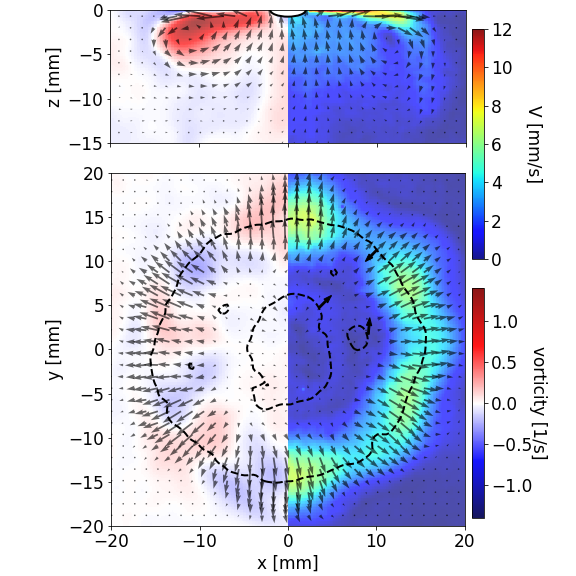}     
       \includegraphics[width=0.49\textwidth]{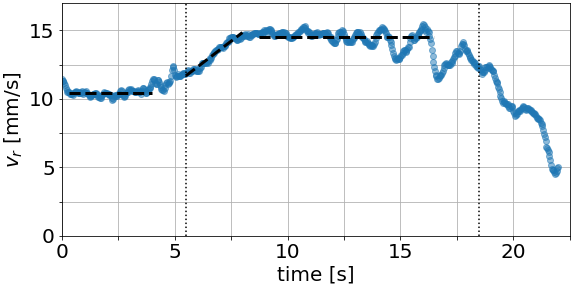}
    \caption{Flow around the spinning drop. $(a)$: vertical cross section in the midplane $y=0$; $(b)$: horizontal cross section in the plane $z=-2.5\,$mm (with $z=0$ corresponding to the highest visible horizontal plane, located slightly below the actual free surface). \textcolor{black}{Snapshots $(a)$ and $(b)$ are taken approximately $4\,$s and $7.7\,$s after the beginning of the spinning, respectively.} The grey arrows show the magnitude and direction of the local in-plane velocity; the colors in the left and right halves refer to the color bars and provide the out-of-plane vorticity component \textcolor{black}{($-\omega_y$ in $(a)$ and $\omega_z$ in $(b)$)} and the magnitude of the in-plane velocity, respectively. In $(b)$, the dashed lines indicate the zero-level of the two-dimensional velocity divergence, the black arrows pointing toward the region of positive values. $(c)$: evolution of the radial velocity, averaged from $r=5\,$mm to $r=15\,$mm in the horizontal plane $z=-0.5\,$mm. The two vertical dotted lines delimit the spinning period.} 
      \vspace{-158mm}\hspace{-66mm}$(a)$\\
      \vspace{49mm}\hspace{-66mm}$(b)$\\
       \vspace{36mm}\hspace{-75mm}$(c)$\\
    \vspace{55mm}
    \label{fig:PIV}
 \end{figure}
 \begin{figure}[!t]
    \centering
         \includegraphics[width=0.49\textwidth]{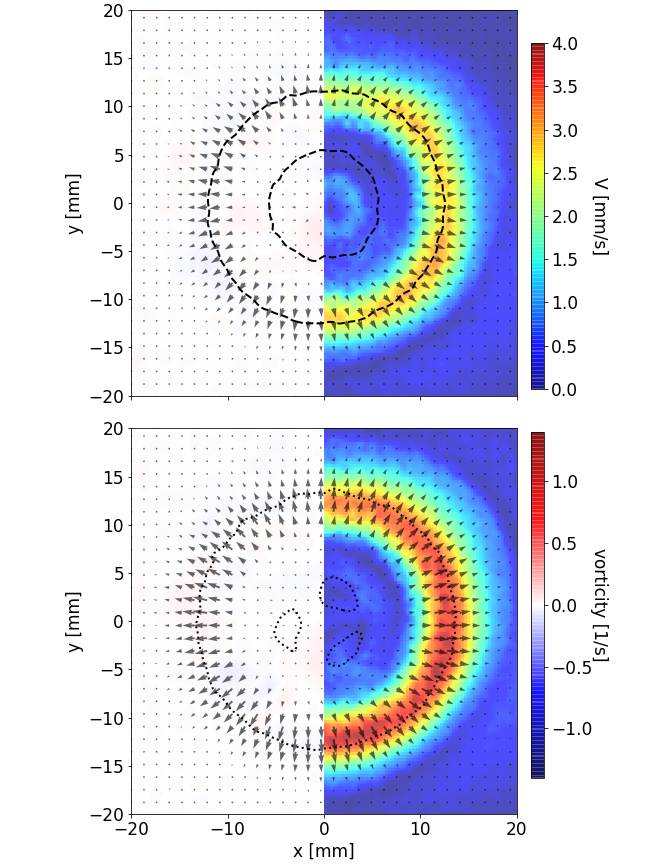}
    \caption{\textcolor{black}{Horizontal cross section of the flow in the plane $z=-2.5\,$mm. $(a)$: during the induction stage, approximately $5\,$s before the beginning of the spinning; $(b)$ approximately $1.3\,$s after the beginning of the spinning. The dashed lines in $(a)$ and the dotted lines in $(b)$ show the iso-contours $\nabla_\parallel\cdot\bm u=0$ and $\nabla_\parallel\cdot\bm u=-0.1$, respectively; for caption, see figure \ref{fig:PIV}.} 
  }
      \vspace{-85mm}\hspace{-63mm}$(a)$\\
      \vspace{48mm}\hspace{-63mm}$(b)$\\
    \vspace{23mm}
    \label{fig:PIV2}
 \end{figure}
Thanks to the PIV setup, the flow inside the aqueous solution can be analyzed all along the drop lifetime. An important outcome of this analysis is that the general flow structure remains unchanged throughout \textcolor{black}{the spinning stage.} 
As depicted in figures~\ref{fig:PIV}$(a)-(b)$, this structure is dominated by an outward Marangoni flow in the near-surface region, \textcolor{black}{driven by the \textcolor{black}{$\Delta\gamma_{w/a}\approx1\,$}mN.m$^{-1}$ surface tension variation between the contact line and the `far field'.} Mass conservation is ensured by a large toroidal recirculation involving an upwelling fluid column below the drop and a slow downward flow taking place far from it. This toroidal  structure is typically $3\,$cm wide and $10\,$mm thick. \textcolor{black}{Note that since the drop does not have any significant vertical motion, the vertical fluid velocity should return to zero in the central part of the plot as the drop surface is approached. However the velocity decrease takes place within a thin near-drop region not covered by the PIV measurements.  \\
\indent The DCM that dissolves in the bath at the contact line and just below it is transported outwards by the radial flow. Its presence at the surface of the bath is evidenced by} a veil visible in the images obtained with the Schlieren technique. \textcolor{black}{Since DCM is significantly denser than water, its dissolution makes the DCM-enriched regions of the bath heavier than the underneath DCM-free water. This is a situation prone to the development of a Rayleigh-Taylor instability in the relevant regions, i.e. below the drop or ahead of the contact line. We never observed any sign of a sinking plume below the drop, nor in the vicinity of the contact line. This leads us to conclude that the Marangoni-driven flow described above completely controls the fluid motion in these regions. In contrast, such plumes may happen well ahead of the contact line, at radial distances where the Marangoni stress and the associated recirculation weaken. This is the case in the right half of figure \ref{fig:PIV}$(a)$, where a downward plume with velocities of the order of $2-3\,$mm.s$^{-1}$ is well visible down to $z\approx-12\,$mm around the radial position $x=15\,$mm. No such structure is observed at the symmetric position $x=-15\,$mm, an indication that slight variations in the flow conditions at a given radial distance (i.e. associated with small circumferential flow asymmetries around the drop, such as those discussed below) may result in stable or unstable conditions with respect to the Rayleigh-Taylor instability.} 
\\ 
\indent \textcolor{black}{To better characterize the flow kinematics, we computed the horizontal velocity divergence and the out-of-plane components of the vorticity. Considering the components $(u_x, u_y, u_z)$ of the local fluid velocity $\bm u$ in the $(x, y, z)$ axes, the $y$- and $z$-components of the vorticity $\boldsymbol\omega=\nabla\times\bm u$ are defined as $\omega_y=\partial_zu_x-\partial_xu_z$ and $\omega_z=\partial_xu_y-\partial_yu_x$, respectively. Not surprisingly, figure \ref{fig:PIV}$(a)$ shows that $\omega_y$ reaches a clear maximum (about $1.2\,$s$^{-1}$) in the `eye' of the recirculation, i.e. at a radial position $r_m\approx11\,$mm and a depth $z_m\approx-2\,$mm. The $\omega_z$-distribution in figure \ref{fig:PIV}$(b)$ reveals an alternation of weak clockwise and anti-clockwise vertical vortices with maximum values in the range $0.2-0.3\,$s$^{-1}$. Six vortex pairs are identified around the drop, each of them being associated with a local maximum of the radial velocity. Neither the position nor the number of these vortices change with the instantaneous drop orientation. Their position may be controlled by the vertical walls of the container (located at $x=\pm25\,$mm). In contrast, the walls play no role in the emergence and selection of this specific pattern, as figure \ref{fig:PIV2}$(a)$ proves. Indeed, this figure shows that the horizontal flow exhibits the expected axial symmetry during the induction stage, with no trace of vertical vortices. Hence, one is led to the conclusion that the initial axisymmetric flow structure loses its symmetry beyond the induction stage and is replaced by the more stable pattern revealed by figure \ref{fig:PIV}$(b)$. \\
\indent Remarkably, the $\omega_z$-distribution shows no trace of the drop rotation. Continuity of horizontal velocities at the drop surface implies $\omega_z=2\Omega$ there, from which values of $\omega_z$ of the order of $20\,$s$^{-1}$ are expected in the drop vicinity. That $\omega_z$ actually takes negligible values at radial distances less than $\approx7\,$mm in the plane $z=-0.5\,$mm suggests that the drop rotation entrains the surrounding fluid only within a thin boundary layer unreachable with the present PIV setup, while leaving the bulk of the bath at rest. This view makes sense since, with $\Omega\approx11\,$rad.s$^{-1}$ and $R_0\approx3.5\,$mm, the Reynolds number based on the spinning rate and physical properties of the aqueous phase ($\mu_w=1\,$mP.s, $\rho_w=1\,$g.cm$^{-3}$) is $Re_\Omega=\rho_wR_0^2\Omega/\mu_w\approx135$, implying that the boundary layer thickness is typically $0.3-0.4\,$mm. Another possibility to reconcile the near-zero values of $\omega_z$ in the central region with the spinning motion revealed by the top views of the system could be that the drop does not actually rotate as a whole, the spinning motion being then only due to a travelling wave deformation along the drop contour. Indeed, it has been shown theoretically \cite{Tarama2012,Tarama2013} that the combination of deformation modes $m=2$ and $m=4$ may lead to a spinning of the contour without any rigid body rotation of the drop. However, we often observed a small blurred region within the drop, corresponding to an emulsion of DCM with the aqueous solution, and could notice that this blurred region was rotating as a whole at the same rate as the tips (see video 2 in SM). We also performed a specific run in which some glass micro-beads were introduced inside the drop. The pattern formed by the micro-beads was seen to rotate with the same rate as the tips, which again supports the view that the drop undergoes a rigid body rotation. 
\\
\indent The horizontal velocity divergence is defined as $\nabla_\parallel\cdot\bm u=\partial_x u_x + \partial_y u_y$. Owing to the incompressibility constraint, one also has $\nabla_\parallel\cdot\bm u=-\partial_z u_z$, which shows that $\nabla_\parallel\cdot\bm u$ gives insight into the vertical variations of the vertical velocity component. In particular, since figure \ref{fig:PIV}$(a)$ indicates that the vertical velocities in the central upwelling region increase as the distance to the surface reduces, $\nabla_\parallel\cdot\bm u$ is expected to be negative in that region. This is confirmed by the shape of the inner $\nabla_\parallel\cdot\bm u=0$-isocontour (figure \ref{fig:PIV}$(b)$) which roughly displays a dilated version of the drop contour. Similarly, in the outer region where the fluid goes downwards, the negative $u_z$ increases with the distance to the surface  in the first few millimeters below it, yielding again $\nabla_\parallel\cdot\bm u<0$. Interestingly, some `islands' in which $\nabla_\parallel\cdot\bm u$ takes negative values may be noticed within the large intermediate region corresponding to positive values. These small regions reveal the upwelling fluid motions underneath the droplets ejected from the tips of the mother drop, since a local flow structure similar to that described above presumably develops around each of these satellite droplets. Similarly, the three `spots' corresponding to $\nabla_\parallel\cdot\bm u<0$ in the central part of figure \ref{fig:PIV2}$(b)$ inform about the three-tip drop geometry at the beginning of the spinning stage.\\ } 
\indent Although the overall flow structure does not change over time, its intensity does, as figure~\ref{fig:PIV}$(c)$ shows. 
Two periods during which this intensity remains nearly constant are observed. During the first of these (corresponding to the lower plateau), the radial velocity is about $10\,$mm.s$^{-1}$. This period covers the end of the induction stage and most of the vibration stage. The second plateau, where the radial velocity is about $14\,$mm.s$^{-1}$ covers most of the spinning stage. Between the two plateaux, the flow accelerates at a rate of $1.3\,$mm.s$^{-2}$. At the end of the spinning stage, the radial velocity exhibits some pulsations before decaying until the drop disappearance.\\
\indent \textcolor{black}{The above magnitude of the radial velocity, $u_r$, may be inferred from simple theoretical arguments \cite{Bandi2017a}. Assuming that the surface tension gradient at the radial position $r$ scales as $\Delta\gamma_{w/a}/r$ and balances the shear stress at the bath surface which scales as $\mu_wu_r(r,z=0)/\delta(r)$ implies that the boundary layer thickness is $\delta(r)\sim\mu_wru_r(r,z=0)/\Delta\gamma_{w/a}$. Since the momentum balance implies $\rho_wu_r^2(r,z=0)/r\sim\mu_wu_r(r,z=0)/\delta^2(r)$, one gets $u_r(r,z=0)\sim\left((\Delta\gamma_{w/a})^2/(\rho_w\mu_wr)\right)^{1/3}$. Averaging from $r=5\,$mm to $r=15\,$mm as in figure \ref{fig:PIV}$(c)$ and considering \textcolor{black}{$\Delta\gamma_{w/a}=1\,$}mN.m$^{-1}$ yields $\overline{u}_r(z=0)\approx4.75\,$cm.s$^{-1}$ and $\overline{\delta}\approx0.45\,$mm, the overbar denoting the radial average. Assuming an approximate exponential decay of the radial velocity across the boundary layer, i.e. $\overline{u}_r(z)/\overline{u}_r(z=0)\approx\exp(-z/\overline{\delta})$ then yields $\overline{u}_r(z=-0.5\text{mm})\approx1.6\,$cm.s$^{-1}$, in good agreement with the plateau value observed in in figure \ref{fig:PIV}$(c)$ during the spinning stage.}

\section{Mechanisms underlying the spontaneous spinning}

We now discuss the mechanisms capable of inducing and sustaining the remarkable spontaneous spinning behavior evidenced in the previous sections. Since the transition to the spinning motion is initiated by the vibrations of the contact line, processes responsible for the latter are examined first. \textcolor{black}{
Then we analyze the mechanisms that make the drop able to develop and sustain a permanent spinning motion once the initial symmetries of the contact line have been broken during the vibration stage.}

\subsection{Vibration triggering}  
Several distinct mechanisms may be at work to trigger the vibrations observed along the contact line. \textcolor{black}{The first of them stands in the variations of the spreading parameter along the induction stage. These variations are induced by the dissolution of DCM, which locally lowers $\gamma_{w/a}$, and the adsorption of CTAB at the drop/solution interface, which lowers $\gamma_{w/o}$.} 
\textcolor{black}{These physico-chemical processes having different characteristic time scales, $S$ varies over time during the induction stage, which triggers oscillations of the contact line.
The second potential source of oscillations resides in the hydrodynamic instabilities that may develop at the drop-bath interface since this interface is sheared by the Marangoni flow, a situation known to allow the development of specific viscous instabilities in two-phase flows. Among them, the best known is Yih's instability \cite{Yih1967}, which is induced by the viscosity jump at the interface. It makes interfacial disturbances grow provided the viscosity of the thickest layer is the lowest. Here the ratio of the bath-to-drop thicknesses is typically $N\approx20:1=20$. The viscosity of DCM under standard conditions being $\mu_o\approx0.4\,$mPa.s, the bath-to-drop viscosity ratio is $M\approx2.5>1$, implying that the viscosity stratification is stabilizing in the present case. 
In contrast, a distinct viscous instability may take place under certain conditions if a surfactant is adsorbed at the same sheared interface \cite{Frenkel2002,Halpern2003}. This instability results from the phase shift between interface deformations and surfactant-induced variations of the interfacial tension along the interface. It makes these deformations grow provided $1<M<N^2$, a condition fulfilled in the present system. Therefore, variations in the CTAB concentration at the drop-bath interface (hence variations in $\gamma_{w/o}$) are expected to trigger the growth of interfacial disturbances.} 
This instability mechanism has already been invoked to explain how droplets are ejected from the pulsating contact line between a floating oil drop and a water bath in the case a soluble surfactant is introduced in the drop ~\cite{Stocker2007}. 
\textcolor{black}{Based on the above analysis, a plausible scenario explaining the triggering of drop vibrations in the present system might be that time variations of the spreading parameter provide oscillations of the drop-bath interface, and these oscillations are amplified by the above surfactant-induced instability.\\ 
\indent However, an alternative and \textit{a priori} much more probable scenario may be drawn by relying on the observation of figures \ref{fig:PIV2} and \ref{fig:PIV}$(b)$. Indeed, these figures indicate that the intensity of the Marangoni flow increases over time until the spinning motion is fully developed. More specifically, the maximum radial velocity observed in the horizontal plane $z=-2.5\,$mm increases from $\approx3\,$mm.s$^{-1}$ at the end of the induction stage (figure \ref{fig:PIV2}$(a)$) to $\approx3.5\,$mm.s$^{-1}$ just after the spinning has started (figure \ref{fig:PIV2}$(b)$), until $\approx7.5\,$mm.s$^{-1}$ in the middle of the spinning stage  (figure \ref{fig:PIV}$(b)$). This gradual increase suggests that the system becomes unstable, i.e. loses its initial axial symmetry, when the Marangoni-driven flow exceeds a critical strength. This threshold may be estimated by considering the ratio of inertial to viscous forces at the contact line. This ratio is characterized by the Reynolds number $Re_0=\rho_wR_0u_r(r=R_0,z=0)/\mu_w$. With the estimate provided in section \ref{Flow} for the characteristic Marangoni-driven radial velocity, one gets $Re_0=\left((\rho_wR_0\Delta\gamma_{w/a})/\mu_w^2\right)^{2/3}$. With \textcolor{black}{$\Delta\gamma_{w/a}=1\,$}mN.m$^{-1}$ and $R_0\approx3.5\,$mm, the Reynolds number at the end of the induction stage is found to be $Re_0\approx230$. Therefore, the emerging scenario is as follows. During the induction stage, $\Delta\gamma_{w/a}$ increases over time, owing to the continual dissolution of DCM, which lowers the surface tension of the bath in the drop surroundings. Beyond a critical point, characterized by a Reynolds number of the order of the above estimate for $Re_0$, the flow along the contact line becomes unstable, making the underlying flow and the drop contour lose their initial axial symmetry. The symmetries present in an axisymmetric flow correspond to those of the $O(2)$ symmetry group. It is known from dynamical systems theory \cite{Golubitsky1988} that the solutions of  systems belonging to this group which emerge through a Hopf bifurcation take the form of rotating waves (RW) or standing waves (SW), the latter resulting from the superimposition of two counterrotating modes with equal magnitudes. The small-amplitude oscillations of the drop contour travelling along the contact line that are observed in the majority of cases ($\approx75\%$ of the runs) during the vibration stage belong to the RW family. Similarly, the stationary three-lobe oscillations observed in approximately $15\%$ of the runs belong to the SW family. In the former case, the primary bifurcation breaks the initial chiral symmetry of the contact line. This is not the case when the bifurcated solution takes the form of a standing wave, and a second bifurcation is then required to break this symmetry. This secondary bifurcation arises when the SW solution has reached a sufficient amplitude for the new base flow to become unstable. According to the observations reported in figure \ref{fig:oscillation_r3}, this secondary bifurcation takes place when the stationary oscillations of the contact line reach a critical amplitude of $\approx4 \%$ of the equilibrium drop radius. }

\subsection{Driving mechanism of the spinning motion}
\begin{figure}[h!]
    \centering
    \includegraphics[width=0.45\textwidth]{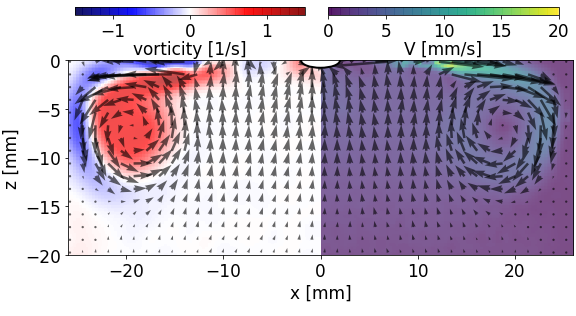}\\
    \vspace{2mm}
    \includegraphics[width=0.45\textwidth]{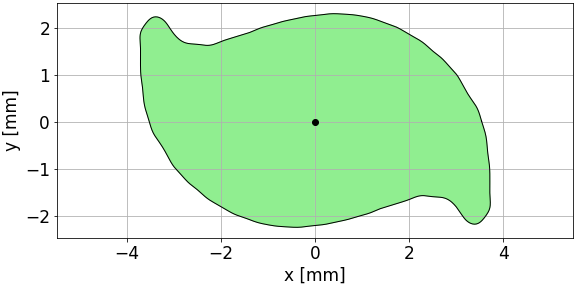}
    \caption{A synthetic `rotor' manufactured with a paper sheet soaked with camphor. $(a)$: vertical cross-section of the flow around the `rotor' (for caption, see figure \ref{fig:PIV}); $(b)$: contour used to produce the `rotor'.   }
       \vspace{-62mm}\hspace{-73mm}$(a)$\\
      \vspace{35mm}\hspace{-73mm}$(b)$\\
    \vspace{19mm}
    \label{fig:side_view_camphor}
   \vspace{-4mm}
\end{figure}

At first glance, three possible mechanisms may be considered as potential candidates to explain the drop spinning once a chiral asymmetry has been introduced along the contact line. These are the repeated ejections of droplets, a hydrodynamic torque resulting from a bathtub-like vortex located below the drop, or a capillary torque induced by the Marangoni effect at the bath surface. However, results presented in the next two subsections allow the first two possibilities to be ruled out easily to the benefit of the latter.\vspace{2mm}\\
\indent\emph{Droplet ejections} -
Since ejected droplets propelling at the surface of the bath are frequently observed during the spinning stage, the mass ejected from the mother drop through these droplets may be suspected to generate a torque that might be responsible for the spontaneous rotation. Such a rocket-like propulsion mechanism (with microbubbles instead of droplets) has already been employed to self-propel microtubular jet engines \cite{Solovev2009}. However, in some runs we also observed that the drop starts spinning without ejecting any droplets, implying that such ejections are not triggering the spontaneous spinning \textcolor{black}{(see video 3 in SM)}. Moreover, droplets are ejected essentially radially (figure \ref{fig:drop}), making them inefficient to generate a vertical torque. 
Last but not least, as figure \ref{fig:spinning_velocity}$(a)$ shows, the spinning rate remains remarkably constant over time, which indicates that the ejection events leave the dynamics of the mother drop unchanged. Based on these observations, we conclude that droplet ejections are a side effect of the spinning, but cannot be its cause.\vspace{2mm}\\
\indent\emph{Hydrodynamic torque} -
As mentioned in section 1, the spontaneous rotation of an ice disk floating on the surface of a water bath has been explained as a consequence of the torque resulting from the bathtub-like vortex created by the sinking into the water mass of the cold water layer produced by the melting ice \cite{Dorbolo2016}. 
In the present system, the dissolution of DCM, whose density is significantly larger than that of water, could generate a qualitatively similar downward plume. 
\textcolor{black}{However, the measurements reported in figure \ref{fig:PIV} prove that the radial Marangoni flow at the surface of the bath dominates the scene, and generates a vigorous upward fluid motion below the drop.} 
These features discard the hydrodynamic mechanism proposed in \cite{Dorbolo2016} as the possible origin of the drop spinning. Moreover, as shown in figure \ref{fig:spinning_velocity}$(d)$, the spinning rate remains unchanged when the depth of the bath is varied. Last, while the radial flow velocity varies by nearly $20\%$ during the spinning stage (figure \ref{fig:PIV}$(c)$), the spinning rate remains remarkably constant. All these observations converge toward the conclusion that, once established, the spinning motion is uncorrelated with the flow. \vspace{2mm}\\
 \indent\emph{Capillary torque} -
A surface tension gradient may induce a capillary torque on a floating object provided the latter exhibits convenient shape asymmetries, or the surface tension differences are suitably distributed around its contour \cite{Pimienta2014}. Since the PIV measurements discussed in section \ref{Flow} evidence the presence of a diverging Marangoni flow at the bath surface, it is tempting to check whether a rigid floating object that modifies the surface tension of the bath in its surroundings and has a contour identical to that of the spinning drop would display the same dynamics.
 To test this hypothesis, we reproduced the drop contour at full scale on a paper sheet soaked with camphor, a substance widely used to produce capillary-driven motions \cite{Ikura2013,Sharma2020}. For this purpose, we recorded a typical two-tip drop shape during the spinning stage, expanded it in Fourier modes and kept only the twenty-four first odd coefficients in this expansion. By doing so, we guaranteed that the contour of the synthetic `rotor' thus obtained has a central symmetry, as figure \ref{fig:side_view_camphor}$(b)$ confirms. We released the `rotor' at the surface of a pure water pool, pinning its centroid to a fixed position but leaving it free to rotate.  A spinning motion immediately set in.  The spinning rate was approximately 2000$^{\circ}$.s$^{-1}$ ($\approx330\,$rpm), more than three times higher than that of the DCM drop. The PIV setup was used to investigate the flow around the `rotor'. As figure \ref{fig:side_view_camphor}$(a)$ shows, the vertical cross section of the flow is qualitatively similar to that reported in figure \ref{fig:PIV}$(a)$, albeit with velocities approximately twice as large. \\
\indent \textcolor{black}{The above side experiment shows that the dynamics and induced flow fields corresponding to the rigid `rotor' soaked with camphor and the DCM drop-CTAB aqueous solution system are extremely similar. This is a strong argument in favor of the Marangoni-induced torque as the driving mechanism responsible for the spinning behavior of the DCM drop. \\
\indent To go one step further into the physical ingredients required to sustain the observed rotation, it is relevant to discuss in some detail the theoretical study of spontaneous translation and rotation of elliptical camphor boats reported in~\cite{Iida2014} (see also \cite{Kitahata2020a}). The model used in that study considers that the translation (resp. rotation) of the boat is driven by a Marangoni force (resp. torque), while the motion is resisted by a viscous force (resp. torque) linearly proportional to the translational (resp. rotational) velocity. The local camphor concentration at the bath surface is governed by a linear diffusion/sublimation equation, and this concentration modifies the surface tension through a nonlinear equation of state. Starting from infinitesimal nonzero translational and rotational velocities of the body, this simple model produces three distinct families of solutions. For large enough translational and rotational resistances, the boat remains at rest. In contrast, it translates steadily broadside on (resp. rotates steadily) when the translational (resp. rotational) viscous resistance is small enough. Therefore this model establishes that, provided the physical control parameters stand in the suitable range, three ingredients suffice to sustain the capillary-induced steady rotation of a floating body: a nonuniform distribution of the surface tension around it, a non-circular (not necessarily asymmetric) body shape, and a nonzero initial angular motion. All three components are present in the case of the DCM drop, the initiation of the angular motion being granted during the vibration stage as discussed above. Interestingly, this model suggests that the asymmetric S-shape taken by the drop is a consequence of its spinning rather than its cause, as only a non-circular contour is required to ensure the body rotation.  A similar conclusion was reached with a purely mechanical system in \cite{Vandenberghe2004}. In that study, a rectangular plate attached to a vertical shaft was oscillated up and down within a circular water tank. Beyond a critical frequency, the flow past the plate is no longer fore-aft symmetric and the plate rotates within the bath, the rotation rate increasing linearly with the forcing frequency.\\
\indent Obviously, inertial effects are too large under present conditions for the model of \cite{Iida2014} to be quantitatively applicable. Indeed, since the Reynolds number based on the rotation rate is beyond $100$ as established in section \ref{Flow}, the resistive viscous torque is actually proportional to $\Omega^2$ rather than $\Omega$ as assumed in \cite{Iida2014}. Nevertheless this leaves the above qualitative conclusions unchanged, and merely modifies the parameter range within which the drop can sustain a steady rotation. It is worth noting that, DCM being denser than water, the drop is thick (see the inset in figure \ref{fig:induction_side}$(c)$). Hence it has to displace a significant amount of fluid in order to translate, i.e. it opposes a large resistance to translation. In contrast, its submerged part being approximately axisymmetric, it only entrains a thin boundary layer when rotating, i.e. it opposes only a small resistance to rotation. Therefore, the present geometrical configuration is favorable to the spinning motion.\\
\indent What is left untouched by the above discussion is the mechanisms that make the drop achieve a constant spinning rate despite its shape and size variations during the spinning stage. Some qualitative arguments help understand this remarkable feature. Considering that the immersed part of the drop looks approximately like a thick disk of characteristic radius $\ell$ and thickness $e$, the time rate-of-change of the drop angular momentum is $T_I=\alpha\rho_oe\ell^4d\Omega/dt$ and the resistive torque is $T_R=-\beta\rho_we\ell^4\Omega^2$, with $\alpha$ a shape-dependent constant and $\beta$ a torque coefficient depending on the spinning Reynolds number. The disk-like shape implies that the pre-factor $\alpha$ weighting the moment of inertia is of order unity. Conversely, the nearly axisymmetric shape of the immersed part of the drop and the large spinning Reynolds number imply that the torque coefficient $\beta$ is much smaller than unity. Therefore the ratio $\alpha/\beta$ is large, indicating that the substantial magnitude of the moment of inertia strongly limits the $\Omega$-variations that could result from an imbalance between $T_R$ and the driving Marangoni torque $T_M$. The latter scales as $(\ell\Delta\ell)\Delta\gamma_{w/a}$, where $\Delta\ell/\ell$ is the characteristic magnitude of the relative non-circularity of the contour shape. Therefore the approximate balance $T_M\approx T_R$ imposed by the above constraints implies that $\rho_w^{-1}(\Delta\ell/\ell)\Delta\gamma_{w/a}\propto e\ell^2$. In other terms, the condition $\Omega^{-2}d\Omega/dt\ll1$ resulting from the combined effect of the drop shape and the large spinning rate forces the product of the relative non-circularity $\Delta\ell/\ell$ and the surface tension variations $\Delta\gamma_{w/a}$ to adjust to the decrease of the drop volume.}

\section{Summary and concluding remarks}

Combining the use of a Schlieren optical technique with two-dimensional PIV measurements, we carried out a detailed experimental investigation of the remarkable dynamics exhibited by a small ($\approx20\,\mu$L) dichloromethane drop placed upon a $10\,$mM CTAB solution. The dynamics of the system exhibit three successive stages. Right after it has been deposited, the drop adopts an axisymmetric lens shape. The magnitude of the (negative) spreading parameter decreases and reaches tiny negative values. This variation is due to the dissolution of dichloromethane near the bath surface on the one hand, and to the adsorption of CTAB at the drop/bath interface on the other hand. Beyond this induction stage which lasts for a few seconds, oscillations occur along the drop contour. Most often, these oscillations take the form of small-amplitude rotating waves corresponding to the azimuthal Fourier mode $m=3$. Stationary oscillations corresponding to the same mode, with an amplitude growing linearly in time, are observed in nearly $15\%$ of the runs. To the best of our knowledge, the spontaneous generation of such stationary odd-mode oscillations with a sessile drop released on a liquid surface has not been reported to date. The stationary pattern becomes unstable when the amplitude of the oscillations exceeds approximately $4\%$ of the drop radius. Whatever the deformation pattern observed during this intermediate step, the initial chiral symmetry of the drop is broken at the end of the vibration stage. Then begins the most spectacular stage during which the drop starts spinning as a rigid body. The drop contour may exhibit two, three or four tips at the beginning of this stage but the last two configurations appear to be unstable and only two-tip contours subsist after a few seconds. Remarkably, the spinning rate stabilizes at approximately $100\,$rpm and no longer varies with the evolution of the drop contour, nor with its gradual shrinking or with the frequent ejection of droplets from the tips. We also checked that changing the depth of the bath or the concentration in CTAB within a significant range leaves the spinning rate unchanged. \\
\indent \textcolor{black}{PIV measurements allowed us to explore the vertical and horizontal structure of the flow in the drop surroundings at various instants of time. Vertical cross sections revealed that} the flow is dominated by a radial outward component near the bath surface. Then, owing to mass conservation, a large toroidal recirculation brings the fluid typically $1\,$cm below the surface, and this fluid returns to the surface in the form of a central upward plume underneath the drop. \textcolor{black}{No trace of the drop spinning is found in the horizontal velocity distributions, suggesting that only a thin boundary layer is entrained by the rotation of the drop/bath interface, a conclusion supported by the large Reynolds number associated with the intense spinning motion.}\\
\indent \textcolor{black}{To build a rational scenario capable of explaining the observed dynamics, especially the spontaneous drop spinning, we explored various mechanisms that can potentially be at work in the system. Based on the Schlieren and PIV measurements, several of them could be easily ruled out, leaving the way clear for a scenario relying entirely on the Marangoni effect at the bath surface.} Indeed, the dissolution dichloromethane in the drop vicinity quickly lowers the surface tension by \textcolor{black}{approximately $1\,$}mN.m$^{-1}$ with respect to the far-field value, generating a radial surface tension gradient directed outwards. \textcolor{black}{Based on the gradual increase of the maximum radial fluid velocity in the drop surroundings, we concluded that the Marangoni-induced flow becomes unstable beyond a critical Reynolds number reached at the end of the induction stage. Given the initial symmetries of the system, rotating waves and standing waves emerge beyond the bifurcation. These bifurcated solutions break the initial rotational symmetry of the system, either directly in the former case or beyond the secondary bifurcation that takes place once the waves have reached a sufficient amplitude in the latter case. The drop contour having lost its initial circular shape and now exhibiting some chiral asymmetry, the system is ready to generate a nonzero Marangoni torque. Following the model developed in \cite{Iida2014}, and noting that the drop geometry induces a large translational resistance but only a small rotational resistance, we concluded that it is then no surprise that the system selects the spinning mode. Noting that the drop geometry results in a substantial moment of inertia while the large spinning rate implies a small resistive torque coefficient, we also provided arguments to explain why the spinning rate stays remarkably constant despite the evolution of the drop shape and size.\\
\indent We plan to make several steps of the above scenario more quantitative by performing dedicated numerical simulations. In particular, the destabilization of the radial Marangoni flow and the flow pattern that emerges from the corresponding primary bifurcation can be studied by solving the three-dimensional Navier-Stokes equations in a single-phase system with a flat surface on which appropriate boundary conditions are imposed. Similarly, starting from an arbitrary initial translational/rotational velocity disturbance, the initiation and development of the spontaneous motion of a drop with a simplified but realistic geometry can be studied numerically in the framework of a low-order model qualitatively similar to that of \cite{Iida2014} but extended to the inertia-dominated regimes relevant to the present experimental conditions. } 

\subsection*{CRediT authorship contribution statement}
\textcolor{black}{{\bf{Dolachai Boniface:}} Investigation, Validation, Formal analysis, Writing - original draft.
{\bf{Julien Sebilleau:}} Conceptualization, Methodology, Formal analysis, Resources, Supervision.
{\bf{Jacques Magnaudet:}} Conceptualization, Formal analysis, Writing - review \& editing, Supervision, Funding acquisition.
{\bf{V\'eronique Pimienta:}} Conceptualization, Methodology, Resources, Supervision, Funding acquisition.}

\section*{Declaration of competing interest}
\textcolor{black}{The authors declare that they have no known competing financial interests or personal relationships that could have appeared to influence the work reported in this paper.}
\section*{Acknowledgements}
\textcolor{black}{We greatly thank S. Cazin and M. Marchal whose expertise with high-speed cameras and PIV was instrumental. The support of the FERMAT Research Federation, which granted D. Boniface's fellowship and provided the cameras used in the experiments, is very much appreciated. This work was partly supported by the Centre National d'Etudes Spatiales (CNES) under grant [APRSDM 6273].}
\section*{Appendix A. Supplementary material}
\textcolor{black}{Supplementary data to this article can be found online at...}

 \bibliographystyle{elsarticle-num} 
\bibliography{biblio_JM}





\end{document}


\title{{\bf{Spontaneous spinning of a dichloromethane drop\\ on \color{black}{an aqueous} \color{black} surfactant solution}}\vspace{2mm}\\
D. Boniface, J. Sebilleau, J. Magnaudet \& V. Pimienta\vspace{3mm}\\{\bf{Supplemental Material: Method details}}} 
\maketitle

\section{Tensiometry}
\begin{figure}[h!]
    \centering
    \includegraphics[width=0.5\textwidth]{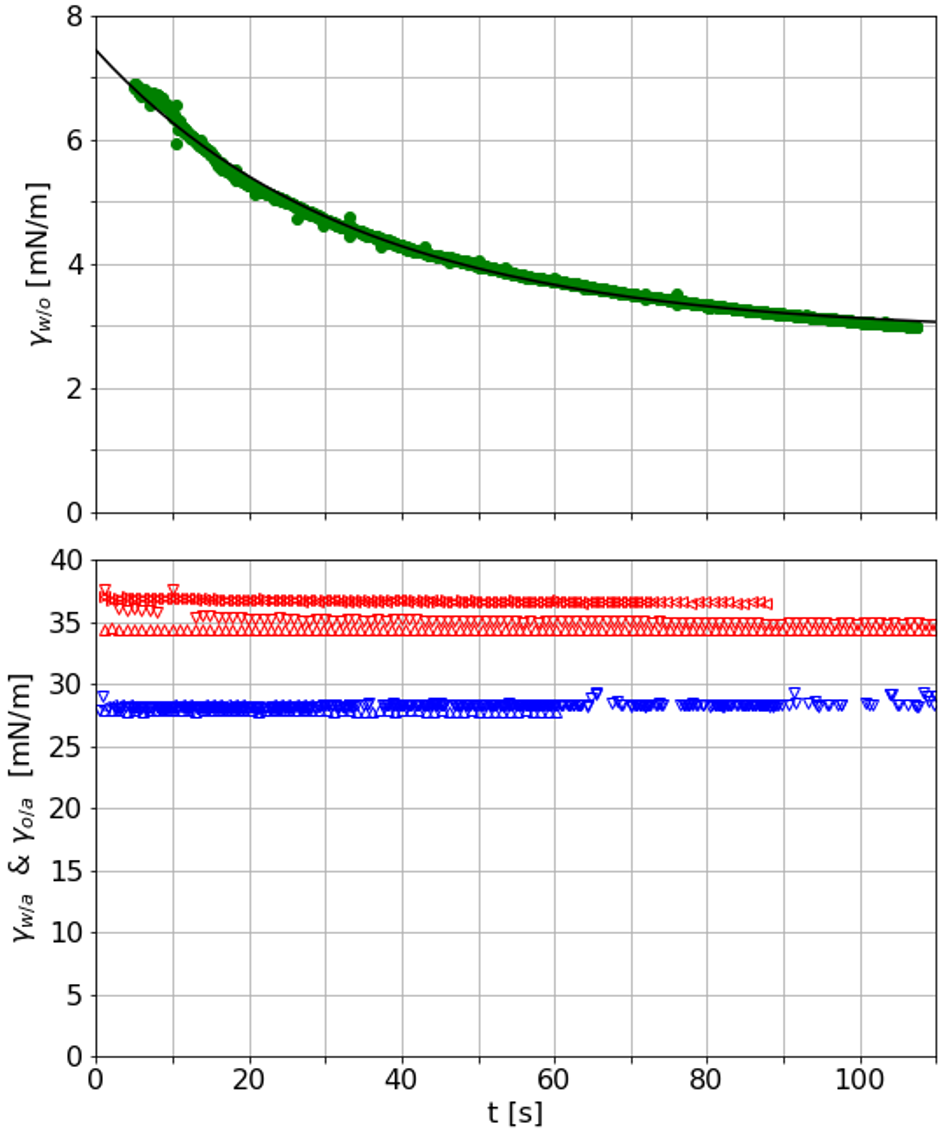}
    \caption{ Time evolution of the three interfacial tensions. $(a)$: $\gamma_{w/o}$ at the DCM /$10\,$mM CTAB solution interface; $(b)$: $\gamma_{w/a}=36\pm1\,$mN.m$^{-1}$ at the $10\,$mM CTAB solution/air interface (red triangles), and $\gamma_{o/a}=28\pm0.3\,$mN.m$^{-1}$ at the DCM/air interface (blue triangles). The black solid line in $(a)$ shows the exponential fit $\gamma_{w/o}=\gamma_\infty+(\gamma_0\textcolor{black}{-}\gamma_\infty)\exp (\textcolor{black}{-}t/\tau_{w/o})$, with $\gamma_0=7\,$mN.m$^{-1}$, $\gamma_\infty=3\,$mN.m$^{-1}$ and $\tau_{w/o}\simeq 30\,$s. In $(b)$, three independent runs are reported for each surface tension. 
    }
       \vspace{-84mm}\hspace{-82mm}$(a)$\\
        \vspace{45mm}\hspace{-82mm}$(b)$\\
            \vspace{35mm}
    \label{fig:tensiometrie}
    \vspace{-3mm}
\end{figure}

\begin{figure}
    \centering
   \hspace{8mm}\includegraphics[width=0.3\textwidth]{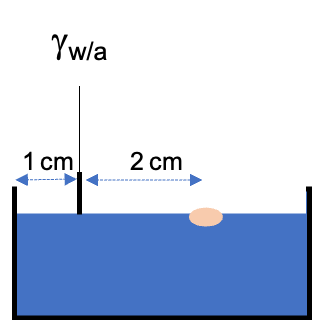}\\
   \includegraphics[width=0.5\textwidth]{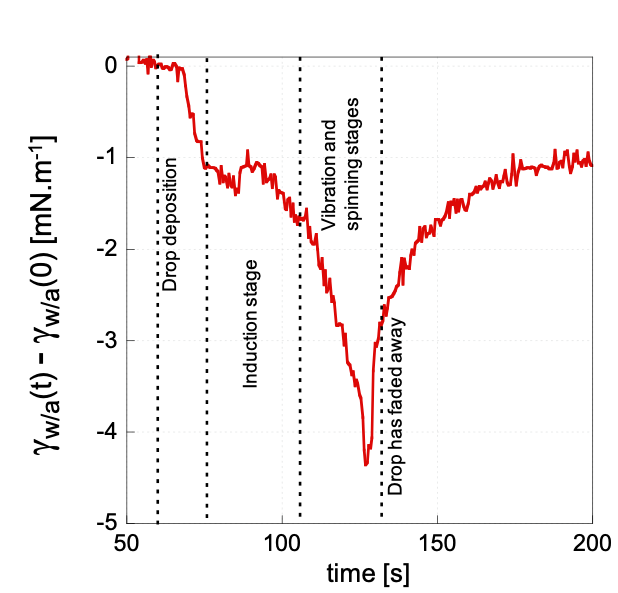}
  \caption{Side experiment revealing the surface activity of DCM on water. $(a)$:
sketch of the measurement setup; $(b)$ typical evolution of the water/air surface tension during and after the deposition of a $20\,\mu$L DCM drop.}
  \vspace{-104mm}\hspace{-82mm}$(a)$\\
        \vspace{72mm}\hspace{-82mm}$(b)$\\
            \vspace{23mm}
      \label{fig:tensiometrie_2}
\end{figure}

\indent
\indent The three interfacial tensions involved in the system were determined with the pendent drop technique using a drop Profile Analysis Tensiometer PAT-1 from SINTERFACE. As may be seen in figure \ref{fig:tensiometrie}, the uncertainty on $\gamma_{w/a}$ is significantly larger than that on $\gamma_{o/a}$, typically $\pm1\,$mN.m$^{-1}$ instead of $\pm0.3\,$mN.m$^{-1}$. Similarly, the uncertainty on $\gamma_{w/o}$ is estimated to $\pm1\,$mN.m$^{-1}$.\vspace{2mm}\\
\indent A separate experiment was carried out to reveal the surface activity of DCM on water. For this purpose, a Wilhelmy plate was inserted $1\,$cm from the wall of the tank and the drop was deposited $2\,$cm away from the plate, i.e. not at the center of the tank since the horizontal section of the latter is $5\,$cm$\times5\,$cm (see figure \ref{fig:tensiometrie_2}$(a)$). A typical evolution of $\gamma_{w/a}$ at the bath surface is displayed in figure \ref{fig:tensiometrie_2}$(b)$. This evolution successively reveals: $(i)$ a fast $\approx1.1\,$mN.m$^{-1}$ decrease of $\gamma_{w/a}$ during the deposition process; $(ii)$ a nonuniform but overall decreasing evolution during the induction stage, decreasing $\gamma_{w/a}$ by an additional $\approx0.6\,$mN.m$^{-1}$; $(iii)$ a sharper decrease during the vibration and spinning stages, culminating in a $\approx4.4\,$mN.m$^{-1}$ reduction of $\gamma_{w/a}$ with respect to its initial value; $(iv)$ a recovery stage during which the surface tension re-increases when the drop fades away; the recovery process goes on beyond  this point but $\gamma_{w/a}$ stays slightly lower than its initial value, due to the contamination of the surface by residual DCM. The evolution shown in figure \ref{fig:tensiometrie_2}$(b)$ is highly reproducible. In particular, compared to its initial value, the surface tension $2\,$cm away from the drop center is found to be always lower by $1.5\pm0.2\,$mN.m$^{-1}$ at the end of the induction stage.

\section{Determination of the spreading parameter}
\indent 
\indent The spreading parameter is defined as $S=\gamma_{w/a}-(\gamma_{o/a}+\gamma_{w/o})$, with all three interfacial tensions evaluated at the contact line. We assume that $\gamma_{o/a}$ and
 $\gamma_{w/o}$ are uniform along the corresponding interfaces and aim at determining $\gamma_{w/a}$ by considering the force balance at the contact line. For this purpose, we introduce the apparent contact angle $\theta_{w/o}$ (resp. $\theta_{o/a}$) between the immersed (resp. emerged) surface of the drop and the horizontal direction, and the meniscus angle $\theta_{w/a}$ between the surface of the bath and the horizontal direction, the inclination of all three surfaces being taken at the contact line. The horizontal and vertical force balances at the contact line then read
 \begin{eqnarray}
 \label{1}
 \gamma_{w/a}\cos\theta_{w/a}= \gamma_{o/a}\cos\theta_{o/a}+ \gamma_{w/o}\cos\theta_{w/o}\,,\\
 \label{2}
  \gamma_{w/o}\sin\theta_{w/o}=\gamma_{o/a}\sin\theta_{o/a}+ \gamma_{w/a}\sin\theta_{w/a}\,.
 \end{eqnarray}
 By squaring (\ref{1}) and (\ref{2}), adding the two and expressing the resulting equation in terms of $S$, one gets 
 \begin{equation}
 S^2+2(\gamma_{o/a}+ \gamma_{w/o})S+4\gamma_{o/a}\gamma_{w/o}\sin^2\frac{\theta_{o/a}+\theta_{w/o}}{2}=0\,.
 \label{3}
 \end{equation}
 Since $S$ is larger than $-(\gamma_{o/a}+ \gamma_{w/o})$ by construction, the only relevant root of (\ref{3}) is
 \begin{equation}
 S=-(\gamma_{o/a}+ \gamma_{w/o})+\bigg\{(\gamma_{o/a}+ \gamma_{w/o})^2-4\gamma_{o/a}\gamma_{w/o}\sin^2\frac{\theta_{o/a}+\theta_{w/o}}{2}\bigg\}^{1/2}\,.
 \label{4}
 \end{equation}
 \indent Experimentally, only $\gamma_{o/a}$, $\gamma_{w/o}$ and $\theta_{w/o}$ are known. In contrast $\theta_{o/a}$ could not be measured with the Schlieren technique. Nevertheless, the upper part of the drop looks very flat (due to the large density of DCM), so that this angle cannot exceed a few degrees. Let us first assume $\theta_{o/a}=0^\circ$. We seek to estimate $S$ at the end of the induction stage, i.e. $\approx10\,$s after the drop has been deposited in the specific case of the drop considered in \S\,3.1 of the paper. With $\gamma_{o/a}=28\,$mN.m$^{-1}$, $\gamma_{w/o}=6\, $mN.m$^{-1}$ (estimated from figure \ref{fig:tensiometrie}$(a)$) and $\theta_{w/o}=23^\circ$, (\ref{4}) implies $S\approx-0.4\ $mN.m$^{-1}$. Assuming now $\theta_{o/a}=6^\circ$, (\ref{4}) implies 
 $S\approx-0.6\ $mN.m$^{-1}$. Thus a median estimate for $S$ is $S=-0.5\pm0.1\ $mN.m$^{-1}$. Varying  $\gamma_{w/o}$ (resp.  $\gamma_{o/a}$) by $\pm1\ $mN.m$^{-1}$ makes $S$ vary only by $\mp0.05\ $mN.m$^{-1}$ (resp. $\mp0.0025\ $mN.m$^{-1}$), while varying $\theta_{w/o}$ by $\pm2^\circ$ changes $S$ by $\mp0.07\ $mN.m$^{-1}$. These various estimates make us confident that the uncertainty on $S$ does not exceed $\pm0.2\,$mN.m$^{-1}$, so that we keep $S=-0.5\pm0.2\ $mN.m$^{-1}$ as the final estimate. From this we deduce $\gamma_{w/a}=33.5\pm1\ $mN.m$^{-1}$, owing to the $\pm1\ $mN.m$^{-1}$ uncertainty on $\gamma_{w/o}$.\\
 \indent Equations (\ref{1}) and (\ref{2}) may also be combined in such a way that $\theta_{o/a}$ is eliminated, allowing $\theta_{w/a}$ to be estimated. The result reads
 \begin{equation}
\cos(\theta_{w/o}-\theta_{w/a})=\frac{\gamma_{w/a}^2+\gamma_{w/o}^2-\gamma_{o/a}^2}{2\gamma_{w/a}\gamma_{w/o}}\,,
 \end{equation}
from which one obtains the estimate $\theta_{w/a}\approx1.5^\circ$.\vspace{3mm}

\section{Influence of PIV tracers on the drop behavior}
\begin{figure}
    \centering
    \vspace{-1mm}
    \includegraphics[width=0.4\textwidth]{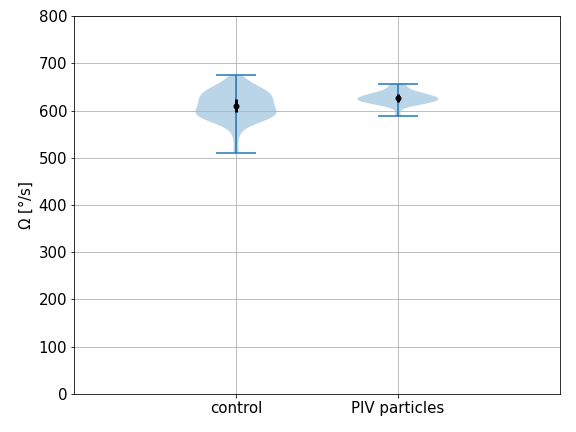}
    \caption{ Influence of PIV tracers on the spinning rate of a DCM drop released on a $2\,$cm deep bath of a $10\,$mM CTAB aqueous solution. The control group corresponds to the reference situation without PIV tracers in the bath. The blue area is the violin representation showing the minimum and maximum values and the probability density of the data; black dots represent the mean value; error bars correspond to a $99\%$ confidence interval obtained from Student's law.}
    \label{fig:tracers}
\end{figure}

\indent
\indent To check the possible influence of PIV tracers (polyamide PA12 $7.25\,\mu$m-diameter spheres containing Rhodamine) on the behavior of the spinning drop, we performed Schlieren measurements with the seeded CTAB aqueous solution and compared them to reference measurements obtained in the absence of the tracers. As shown in figure \ref{fig:tracers}, this comparison did not reveal any significant influence of the tracers.\\
\newpage
\section{Videos}
\noindent Video 1: Complete evolution of a DCM drop (with a $20\,\mu$L initial volume) spinning on a $2\,$cm deep bath of a $10\,$mM CTAB aqueous solution, from its deposition to its final disappearance.   The movie is displayed at real speed.\vspace{3mm}\\
Video 2: A DCM drop (with a $20\,\mu$L initial volume) spinning on a $2\,$cm deep bath of a $10\,$mM CTAB aqueous solution. Within the drop, an emulsion between the dichloromethane and the aqueous solution has formed spontaneously in a small region. Following the two spots in which this emulsion concentrates allows us to confirm that the drop rotates as a rigid body. The contrast has been modified to better visualize the two spots and the movie has been slowed down seven times.\vspace{3mm}\\
Video 3: A DCM drop (with a $20\,\mu$L initial volume) spinning on a $2\,$cm deep bath of a $10\,$mM CTAB aqueous solution. The movie shows the most common behavior observed during the vibration stage, with oscillations of the contact line propagating along it, and the early spinning stage which in that case involves a two-tip spinner. Here the spinning behavior is initiated without any droplet emission from the tips. The movie is slowed down two and a half times.\vspace{3mm}\\
Video 4: A DCM drop (with a $20\,\mu$L initial volume) spinning on a $2\,$cm deep bath of a $10\,$mM CTAB aqueous solution. The movie shows the vibration stage with the emergence of stationary oscillations of the contact line corresponding to the mode $m=3$, followed by an early spinning stage which in that case involves a three-tip spinner. The movie is slowed down five times.\vspace{3mm}\\
Video 5: A DCM drop (with a $20\,\mu$L initial volume) spinning on a $2\,$cm deep bath of a $10\,$mM CTAB aqueous solution. The movie shows a `chaotic' vibration stage followed by the (rare) emergence of a four-tip spinner. The movie is slowed down two and a half times.

%% file: macros.tex
\definecolor{mygray}{gray}{0.4}